\documentclass[pra,twocolumn,superscriptaddress,10pt,floatfix]{revtex4-1}
\usepackage[english]{babel}
\usepackage{amsmath,amssymb,bm,bbm}
\usepackage{graphicx,xcolor,xspace,hyperref}
\usepackage[normalem]{ulem}
\usepackage{cancel}
\xspaceaddexceptions{]\}+}

\newcommand{\wt}[1]{\widetilde{#1}}

\newcommand{\ket}[1]{{\lvert#1\rangle}}
\newcommand{\bra}[1]{{\langle#1\rvert}}

\newcommand{\lrangle}[1]{\langle#1\rangle}

\newcommand{\R}{\mathbb{R}}
\newcommand{\Z}{\mathbb{Z}}
\newcommand{\A}{\mathcal{A}}
\newcommand{\B}{\mathcal{B}}
\newcommand{\N}{\mathcal{N}}
\newcommand{\CC}{\mathcal{C}}
\newcommand{\II}{\mathcal{I}}
\newcommand{\XX}{\wt{X}}
\newcommand{\ZZ}{\wt{Z}}
\newcommand{\q}{\wt{q}}
\newcommand{\LL}{\mathcal{L}}

\newcommand{\Hc}{\mathrm{H.c.}}
\newcommand{\Tr}{\mathrm{Tr}}

\newcommand{\sep}{\ensuremath{\mathrm{sep}}}
\newcommand{\ns}{\ensuremath{\mathrm{ns}}}
\newcommand{\Potts}{\ensuremath{\mathrm{Potts}}}
\newcommand{\even}{\ensuremath{\mathrm{even}}}
\newcommand{\odd}{\ensuremath{\mathrm{odd}}}

\newcommand{\U}[1]{\ensuremath{\operatorname{U(}\!#1\!\operatorname{)}}\xspace}
\newcommand{\SU}[1]{\ensuremath{\operatorname{SU(}\!#1\!\operatorname{)}}\xspace}
\newcommand{\SO}[1]{\ensuremath{\operatorname{SO(}\!#1\!\operatorname{)}}\xspace}

\newcommand{\xFM}{\ensuremath{{x\!\operatorname{FM}}}\xspace}
\newcommand{\zFM}{\ensuremath{{z\!\operatorname{FM}}}\xspace}
\newcommand{\VBS}{\ensuremath{{\operatorname{VBS}}}\xspace}

\newcommand{\dwt}[1]{\widetilde{\raisebox{0pt}[0.89\height]{$\widetilde{#1\mkern0mu}$}}}

\newcommand{\ee}{\mathrm{e}} 
\newcommand{\ii}{\mathrm{i}\,} 

\begin{document}

\title{One-dimensional model for deconfined criticality with \texorpdfstring{$\Z_3 \times \Z_3$}{Z3xZ3} symmetry}

\date{\today}

\author{Brenden Roberts}
\email{broberts@caltech.edu}
\affiliation{Institute for Quantum Information and Matter,\\California Institute of Technology, Pasadena, CA 91125}
\author{Shenghan Jiang}
\email{jiangsh@ucas.ac.cn}
\affiliation{Institute for Quantum Information and Matter,\\California Institute of Technology, Pasadena, CA 91125}
\affiliation{Kavli Institute for Theoretical Sciences and CAS Center for Excellence in Topological Quantum Computation, University of Chinese Academy of Sciences, Beijing 100190, China}
\author{Olexei I.~Motrunich}
\email{motrunch@caltech.edu}
\affiliation{Institute for Quantum Information and Matter,\\California Institute of Technology, Pasadena, CA 91125}

\begin{abstract}
We continue recent efforts to discover examples of deconfined quantum criticality in one-dimensional models.
In this work we investigate the transition between a $\Z_3$ ferromagnet and a phase with valence bond solid (VBS) order in a spin chain with $\Z_3\times\Z_3$ global symmetry.
We study a model with alternating projective representations on the sites of the two sublattices, allowing the Hamiltonian to connect to an exactly solvable point having VBS order with the character of \SU3-invariant singlets.
Such a model does not admit a Lieb--Schultz--Mattis theorem typical of systems realizing deconfined critical points. 
Nevertheless, we find evidence for a direct transition from the VBS phase to a $\Z_3$ ferromagnet.
Finite-entanglement scaling data are consistent with a second-order or weakly first-order transition.
We find in our parameter space an integrable lattice model apparently describing the phase transition, with a very long, finite, correlation length of 190878 lattice spacings.
Based on exact results for this model, we propose that the transition is extremely weakly first order, and is part of a family of DQCP described by walking of renormalization group flows.
\end{abstract}

\maketitle
\tableofcontents

\section{Introduction}

One of the broad objectives of recent study in condensed matter physics is to describe quantum phase transitions outside the scope of the usual Landau--Ginzburg theory of symmetry breaking.
Within this topic, a number of spiritually similar proposals fall into the category of ``deconfined quantum critical points'' (DQCP).
This label was originally used for a model of spins with \SU2 symmetry on the two-dimensional square lattice to describe a transition between a phase with N{\'e}el antiferromagnetic order and a one with columnar valence-bond solid (VBS) order.
\citet{DQCP_science,DQCP_prb} proposed a mechanism for a continuous transition which relies on emergent symmetry, leading to a theory in terms of fractionalized fields.
This description inspired a variety of other proposals, which are united by the property that the natural variables for the system at the critical point are confined---and thus absent at low energies---in either phase.
Meanwhile, the original proposal has been extensively tested in numerical studies, which are consistent with either a second-order or very weakly first-order transition \cite{Sandvik2007, MelkoKaul2008, LuoSandvikKawashima2009, BanerjeeDamleAlet2010, Sandvik2010, HaradaSuzukiOkuboMatsuoLuoWatanabeTodoKawashima2013, JiangNyfelerChandrasekharanWises2008, ChenHuangDengKuklovProkofevSvistunov2013, NahumChalkerSernaOrtunoSomoza2015, NahumSernaChalkerOrtunoSomoza2015II, MotrunichVishwanath2008, KuklovMatsumotoProkofevSvistunovTroyer2008, Bartosch2013, CharrierAletPujol2008, ChenGukelbergerTrebstFabienBalents2009, CharrierAlet2010, SreejithPowell2015, ShaoGuoSandvik2016}.

The low-energy theory for the N{\'e}el-VBS transition is the non-compact CP$^1$ model describing complex scalars coupled to a \U1 gauge field which however does not include monopole terms in the action.
Quantum Monte Carlo simulations suggest that the IR theory of the NCCP$^1$ model hosts an emergent symmetry, with the three components of the N{\'e}el order parameter and two components of the VBS order parameter transforming together as an \SO5 vector \cite{nahum2015emergent}.
This emergent symmetry, which is realized anomalously, proved to be useful for developing an understanding of the transition through various dualities to theories which can appear on the surface of a three-dimensional symmetry protected topological (SPT) phase \cite{wang2017deconfined}.

Surprisingly, conformal bootstrap bounds on unitary CFTs with SO(5) symmetry turn out to exclude the conformal data measured in numerics, including for the \SO5 vector which is too relevant to satisfy consistency conditions.
The resolution may be that the phase transition is in fact weakly first order (pseudo-critical), a phenomenon thought to be generically a result of renormalization group walking.
In this scenario, the transition displays approximate conformal symmetry below some long, but finite, length scale.
At intermediate distances the system's properties are governed by non-unitary complex fixed points which can be viewed as analytic continuations of a unitary CFT; however eventually the theory becomes gapped.
For the DQCP with \SU2 symmetry such a description requires a fixed point with inherent \SO5 symmetry and a tunable parameter providing access to the pseudo-critical regime \cite{wang2017deconfined}.
Some proposals in this direction have identified as a candidate a nonlinear sigma model with WZW term continued to $d = 2+\epsilon$ dimensions with $\SO{4+\epsilon}$ symmetry  \cite{nahum2019note,ma2020theory}.

A complementary perspective on the above story arises from framing the phenomenology of the DQCP in models in one dimension, where one breaks the global symmetry to some discrete subgroup.
In Ref.~\cite{jiang2019ising} a transition was considered between a ferromagnet and a dimerized \VBS phase in a one-dimensional system with $\Z_2\times\Z_2$ symmetry.
Exact lattice dualities lead to a mapping to microscopic variables which unify these order parameters and allow a controlled low-energy theory, which turns out to be a Luttinger liquid with a single relevant cosine term and continuously varying critical indices.
In these deconfined variables an emergent $\U1\times\U1$ symmetry is manifest at the transition.
Studies of a concrete spin system established many nontrivial properties of this theory \cite{roberts2019deconfined,huang2019emergent,mudry2019quantum}.
Another example of DQCP in one dimension has also been observed by using long-ranged Heisenberg terms to circumvent the Mermin--Wagner theorem; such a model (which can be realized on the boundary of a SPT state in two dimensions \cite{jian2020continuous}) exhibits a direct transition between a gapless phase with AFM order and one with \VBS order \cite{yang2020deconfined}.

One may wonder to what extent the lessons learned from the $\Z_2\times\Z_2$-symmetric DQCP in one dimension are representative of a more general class, as opposed to being somehow special.
In the present work we begin to address this question through detailed studies of a concrete lattice model with $\Z_3\times\Z_3$ symmetry.
We will end up arguing that the evidence suggests that a family of DQCP in $\Z_q\times\Z_q$-symmetric models in one dimension in fact exhibits pseudo-critical behavior due to walking, similar to the current status of the canonical DQCP with \SU2 symmetry in two dimensions.
The putative transition in our $\Z_q\times\Z_q$-symmetric DQCP appears to be described by an integrable model with very long correlation length, and the availability of analytical results make it a particularly appealing case for controlled studies of the RG walking scenario for a very weakly first-order DQCP.

This paper is organized as follows.
In Secs.~\ref{sec:Z3xZ3} and \ref{sec:vumps} we introduce our lattice Hamiltonian and present numerical results from matrix product states on the phase diagram and evidence for a DQCP.
In Sec.~\ref{sec:theories} we present some low-energy continuum pictures related to the lattice model and calculate supporting results in a Gaussian theory.
In Sec.~\ref{sec:integrable} we provide details on exact results for an integrable model suggested by numerics to describe the DQCP, which leads us to conclude the transition is the weakly first order.
In Sec.~\ref{sec:ed} we use exact diagonalization studies to identify some light primary fields in the complex CFTs associated with this model.
Finally, in the appendices we expand on background information and further technical details related to various aspects of this work.

\section{Model with \texorpdfstring{$\Z_3 \times \Z_3$}{Z3xZ3} symmetry}
\label{sec:Z3xZ3}

A quantum chain respecting an internal $\Z_3 \times \Z_3$ symmetry is most naturally realized using a three-dimensional local Hilbert space, placed on the sites of a 1d lattice.
We provide detailed motivation and clarification about the form of our Hamiltonian by reviewing the group \SU3 and relevant previous results on lattice models with \SU3 symmetry in App.~\ref{app:su3}.

\subsection{Lattice Hamiltonian}

We choose the following generators of the global internal symmetry group:
\begin{equation}
g_x = \prod_j g_{x,j} = \prod_j X_j,~g_z = \prod_j g_{z,j} = \prod_k Z^\dag_{2k} Z_{2k+1},
\end{equation}
which are written using the $\Z_3$ clock operators
\begin{equation}
X = \begin{bmatrix}0&0&1\\1&0&0\\0&1&0\end{bmatrix},~~Z = \begin{bmatrix}1&0&0\\0&\omega&0\\0&0&\omega^{-1}\end{bmatrix},
\end{equation}
with $\omega = e^{\ii 2 \pi /3}$ being the primitive cubic root of unity.
Because of the commutation relation $ZX = \omega XZ$ the $\Z_3^z \times \Z_3^x$ symmetry is realized projectively on a single lattice site.
The projective representations are classified by \mbox{$H^2[\Z_3 \times\Z_3,\U1] = \Z_3$} and labeled by $\{[0],[1],[2]\}$, where for class $[r]$ we have $g_{z,j} g_{x,j} = \omega^r g_{x,j} g_{z,j}$. 
The sublattice of odd-numbered (even-numbered) sites hosts the $[1]$ ($[2]$) projective representation of $\Z_3 \times \Z_3$.

The general lattice Hamiltonian we consider is
\begin{widetext}
\begin{align}
H = H[J^x,J^z,K] &= - \sum_j \left( ( J^x X_j X_{j+1} + J^z Z^\dag_j Z_{j+1} + \Hc ) + K (1 + X_j X_{j+1} + \Hc)(1 + Z^\dag_j Z_{j+1} + \Hc) \right) \label{eq:Hz3}\\
&= - \sum_j \left( J^x X_j X_{j+1} + J^z Z^\dag_j Z_{j+1} + \Hc \right) + 6 K \sum_j \left( \sum_a \overline T^a_j T^a_{j+1} - \frac{1}{6}\right) ~.
\end{align}
\end{widetext}
In the second line the $K$ term is written using standard $\SU3$ spin operators connecting to an integrable model with VBS ground state, as reviewed in App.~\ref{app:su3}.
We generally restrict all coupling constants to be real and non-negative.

Other internal symmetries of Eq.~\eqref{eq:Hz3} include time reversal $\Theta$, which we implement as complex conjugation in the $Z$ eigenbasis, and charge conjugation symmetry $\CC:\ket{n} \to \ket{3-n \mod 3}$.
Together $\CC$ and $g_x$ generate the $S_3$ permutation symmetry of the local basis state labels.
With periodic boundaries on the lattice, the model is invariant under the generator of translation $T_1$, as well as spatial inversion $\II$ about a site.
While $T_1$ is a symmetry of $H$, it does exchange the projective symmetry groups on the sublattices.
The action of the symmetries on the clock operators is
\begin{align}
g_x:~~& (X_j,Z_j)~\mapsto~(X_j,\omega^{-1} Z_j)~,\\
g_z:~~& (X_j,Z_j)~\mapsto~(\omega^{2 p_j-1} X_j,Z_j)~,\\
\Theta:~~& (X_j,Z_j)~\mapsto~(X_j,Z^\dag_j),~\ii \mapsto -\ii~,\\
\CC:~~& (X_j,Z_j)~\mapsto~(X^\dag_j,Z^\dag_j)~,\\
T_1:~~& (X_j,Z_j)~\mapsto~(X_{j+1},Z_{j+1})~,\\
\II:~~& (X_j,Z_j)~\mapsto~(X_{-j},Z_{-j})~.
\end{align}
We use $p_j$ to denote the parity of $j$:
\begin{equation}
p_j = \frac{1 - (-1)^j}{2} = \begin{cases}
0~, ~~j~\text{even}~, \\
1~, ~~j~\text{odd}~.
\end{cases}
\label{eq:p_j}
\end{equation}
The relation defining the nontrivial projective representation on site $j$ is
\begin{equation}
g_{x,j} g_{z,j} = \omega^{1 - 2 p_j} g_{z,j} g_{x,j}~.
\end{equation}

\subsection{Classical picture of phases}

In the limiting case $J^x = K = 0$, $J^z > 0$, the ground state is a ferromagnetic phase in the $Z$ basis which breaks $\Z_3^x$, leading to a three-dimensional ground state manifold spanned by basis
\begin{equation}
\B_\zFM = \left\{\bigotimes_j \ket{0}_j,~\bigotimes_j \ket{1}_j,~\bigotimes_j \ket{2}_j\right\}.
\end{equation}
The ground states in the \zFM phase are of course subject to quantum fluctuations but remain connected to this simple basis of product states.

Similarly, for $J^z = K = 0$, $J^x > 0$ the ground states exhibit ferromagnetic order in the $X$ eigenbasis (local basis states denoted $\ket{0_x}$, $\ket{1_x}$, $\ket{2_x}=\ket{-1_x}$):
\begin{equation}
\B_\xFM = \left\{\bigotimes_j \ket{0_x}_j,\bigotimes_j \ket{(1\!-\!2 p_j)_x}_j,\bigotimes_j \ket{(2 p_j\!-\!1)_x}_j\right\}.
\label{eq:BxFM}
\end{equation}

Setting $J^z = J^x = 0$, $K > 0$ recovers the Hamiltonian $H_\text{bQ}$ of Eq.~\eqref{eq:Hbq} which respects the full \SU3 symmetry.
As described in Sec.~\ref{sec:su3hamiltonians}, the ground state of this model is known to preserve $\SU3$ but spontaneously breaks the translation symmetry generator $T_1$ to $T_2 = (T_1)^2$, thus breaking a $\Z/2\Z = \Z_2$ symmetry and leading to twofold ground state degeneracy \cite{affleck1990exact}.
While the ground states at this point are finitely correlated, including additional terms discussed in Sec.~\ref{sec:su3hamiltonians} connects to a Majumdar--Ghosh-like point in the same phase.
Thus we take the classical picture of the \VBS phase to be spanned by
\begin{equation}
\B_\VBS = \left\{\bigotimes_k \ket{\psi_\mathrm{s}}_{2k-1,2k},\bigotimes_k \ket{\psi_\mathrm{s}}_{2k,2k+1}\right\},
\end{equation}
where $\ket{\psi_\mathrm{s}}_{j,j'} = \frac{1}{\sqrt{3}} \left(\ket{00}_{j,j'} + \ket{11}_{j,j'} + \ket{22}_{j,j'} \right)$.

Although every unit cell hosts a nontrivial projective representation, this system does not have an LSM anomaly \cite{SongFangQi2020,else2020topological,JiangChengQiLu2019}, and it turns out that one can construct a gapped symmetric ground state.
This symmetric phase is actually an SPT phase characterized by a fractionalized entanglement spectrum; as such, there is no simple classical picture of this state.
In App.~\ref{app:symm_mps} we develop an analytic MPS for this phase.

\section{Results from uniform matrix product states} \label{sec:vumps}

In order to reduce the three-dimensional parameter space of Eq.~\eqref{eq:Hz3} to a two-dimensional phase diagram, we perform a change of variables to the anisotropy $\delta = \frac{J^z-J^x}{J^z + J^x}$; that is, $J^z = J(1+\delta)$ and $J^x = J(1-\delta)$, and we set $J = 1$.
We find the phase diagram using the variational uniform matrix product state numerical method \cite{zauner2018variational}.
We use an adiabatic protocol for determining the phase boundary, fully optimizing a trial state far away from the transition, then using this trial state as the initial condition for the variational procedure with a slightly perturbed Hamiltonian.
In this way the state is tuned towards the phase transition but biased towards a particular symmetry-breaking order.
Because at the mean-field level the phase transition is first-order, the energy landscape of the MPS close to the transition will develop two local minima, with one being metastable on each side.
As the two choices of initial conditions locate the trial states close to one or the other energy minimum, a comparison of trial energies allows us to determine very precisely the exact location of the crossing for a given bond dimension.
Then finite-entanglement scaling with bond dimension provides an estimate of the true location of the phase transition, based on the understanding of MPS as a dressed mean-field approximation \cite{liu10symmetry}.

For the purposes of uniformity, we add a very small symmetry-breaking term to the Hamiltonian when preparing initial variational states, so that all data are comparable across values of $\chi$.
In particular, in the state coming from the \zFM side, we break $g_x$ by biasing toward $\otimes_j \ket{0}_j$, as this ground state is invariant under the $\CC$ symmetry generator.
All scans are performed independently of one another.

\subsection{Numerical phase diagram} \label{sec:pd}

\begin{figure}[ht]
\includegraphics[width=\columnwidth]{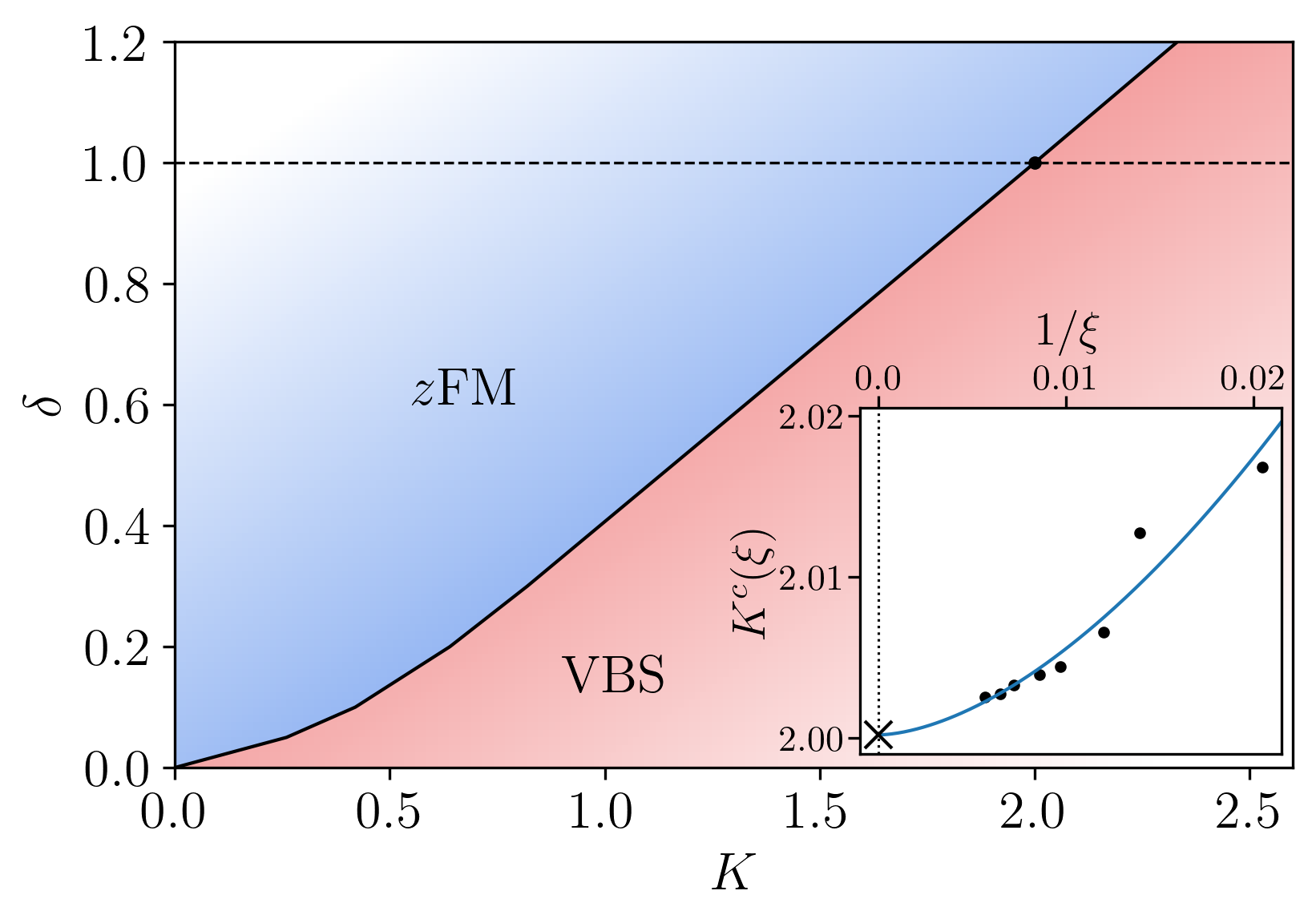}
\caption{\label{fig:pd} The phase diagram of $H[\delta,K]$ is determined from extrapolation in MPS correlation length of optimized variational MPS using an adiabatic protocol.
The dashed line at $\delta=1$ has an enhanced $\U1\times\U1$ onsite symmetry.
The inset shows an example of the finite-entanglement process of approximating $K^c$.
Each data point indicates a crossing of trial energies for states biased towards each symmetry-breaking order, which we scan along slices of constant $\delta$.
The data shown is for $\delta=1$, with bond dimensions from 90 to 300 and correlation lengths between roughly 50 and 175 lattice spacings.
The numerically extrapolated critical point is $K_c(\xi \to \infty) = 2.0002$.
We examine the $(\delta,K)=(1,2)$ point in the phase diagram in detail in Secs.~\ref{sec:integrable} and \ref{sec:ed}.
}
\end{figure}

As we will describe in Sec.~\ref{sec:duality}, the point $(\delta,K) = (0,0)$ maps under duality to two decoupled three-state clock models tuned to the self-dual point, supported on the two sublattices of the dual lattice.
The critical theory describing each sublattice is the CFT for the three-state self-dual Potts model, the minimal model with $c=4/5$.
Accordingly, this point in the phase diagram is critical with $c=8/5$.
The $K$ perturbation in this language has the form of an energy-energy term coupling the two clock models in a way that preserves self-duality.
The corresponding field theory operator is RG relevant but is in fact integrable, known to lead to a massive fixed point~\cite{leclair1998integrability} which presumably describes the \VBS phase in our context.
The $\delta$ term has support on the energy operator---for each of the two Potts models---and is strongly relevant, breaking self-duality and prohibiting a convincing perturbative expansion about the field theory at this point.
(It is interesting that the model with only $\delta$ perturbation is also an integrable deformation of this CFT \cite{zamolodchikov1988integrals}.)

The numerical data are consistent with a ``wedge'' shape; that is, at $\delta=0$ the system is in the \VBS phase for any finite $K > 0$.
The shape of the phase boundary is shown in Fig.~\ref{fig:pd}, where the location of the transition is determined by adiabatic scans (described in the previous section) along cuts of fixed $\delta$, and we extrapolate $\chi \to \infty$.
Away from $\delta=0$ the numerical data are consistent with the conclusion that the transition between \zFM and \VBS ordered phases is second-order, without continuously varying critical exponents.
However, as we describe later, the situation turns out to be more complicated.

The slice $\delta=1$ is indicated on Fig.~\ref{fig:pd}, which in the original parameters of Eq.~\eqref{eq:Hz3} sets $J^x = 0$ and $J^z = 2$.
For $J^x = 0$ the Hamiltonian takes a simpler form:
\begin{align}
H[&J^x=0,J^z,K] \nonumber \\
&= -3\sum_j \Big( J^z\sum_\alpha \ket{\alpha \alpha}\! \bra{\alpha\alpha}_{j,j+1} + K \sum_{\alpha,\beta}\ket{\alpha \alpha}\! \bra{\beta \beta}_{j,j+1}\nonumber \\
&\qquad\qquad\qquad- (J^z + K) \Big)~.
\label{eq:Hu1}
\end{align}
Along this line the global symmetry $\Z_3^z \times \Z_3^x$ is enhanced to $\U1^2 \rtimes~ \Z_3^x$, where generators of the $\U1\times\U1$ symmetry can be constructed from any independent linear combinations of $Z$ and $Z^\dag$ 
\footnote{That is, the $\U1^2$ contains rotations about the generators of the Cartan subalgebra of $\mathfrak{su}(3)$.
In general, in this way a $q$-state model can be written which is symmetric under a $\U1^{q-1}$ subgroup of \SU{q}.}.

We represent the $\U1\times\U1$ symmetry generators by
\begin{align}
\N_1 = \sum_j n_{1,j} = \sum_j (-1)^j \ket 1 \bra 1_j~, \label{eq:n1def} \\
\N_2 = \sum_j n_{2,j} = \sum_j (-1)^j \ket 2 \bra 2_j~. \label{eq:n2def}
\end{align}
A group element is written
\begin{equation}
u(\varphi_1,\varphi_2) = \prod_j e^{\ii (\varphi_1 n_{1,j} + \varphi_2 n_{2,j})}~,
\end{equation}
and we have $g_z=u(-2\pi/3,2\pi/3)$.
The action of the other symmetry generators on $n_{a,j}~(a=1,2)$ is given by
\begin{align}
g_x:~~& n_{1,j} \mapsto n_{2,j},~ n_{2,j} \mapsto (-1)^{j}-n_{1,j}-n_{2,j}~, \label{eq:U1xU1_gx}\\
\Theta:~~& n_{a,j} \mapsto n_{a,j},~ \ii \mapsto -\ii~, \\
\CC:~~& n_{1,j} \mapsto n_{2,j},~ n_{2,j} \mapsto n_{1,j}~, \\
T_1:~~& n_{a,j} \mapsto -n_{a,j+1}~, \\
\II:~~& n_{a,j} \mapsto n_{a,N-j}~.
\label{eq:commutator_n1n2}
\end{align}
Note that the appearance of $(-1)^j$ in Eq.~\eqref{eq:U1xU1_gx} indicates that each site forms a projective representation of the onsite symmetry group generated by $g_x$ and $\N_{1,2}$.
Furthermore, $g_x$ commutes with $\N_{1,2}$ only in the $\N_1 = \N_2 = 0$ sector.

\subsection{Central charge}

\begin{figure}[ht]
\includegraphics[width=\columnwidth]{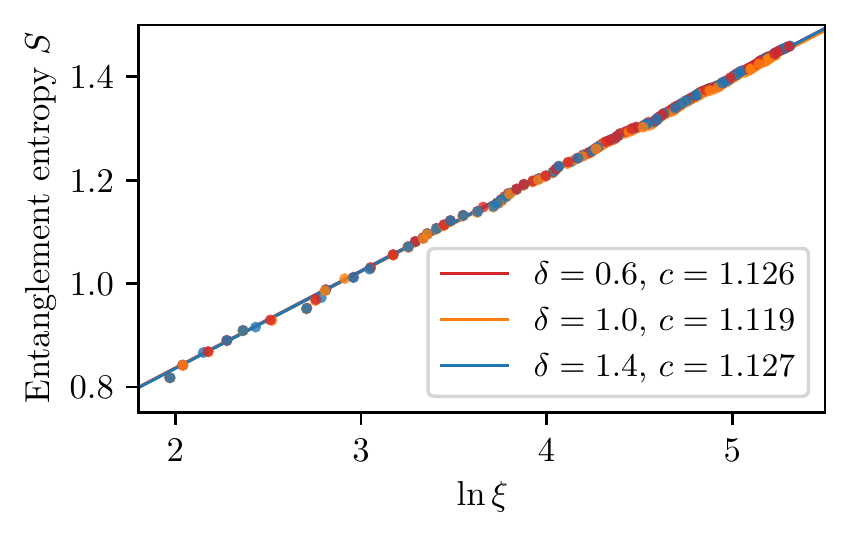}
\caption{\label{fig:central_charge_delta}
Entanglement scaling is shown at the precise phase transition for several values of $\delta$.
We draw data points in random order to emphasize consistency.
Numerical $c$ are obtained by fits to critical scaling of entanglement entropy $S = \frac c6 \ln \xi$.
States are optimized at the critical point but break $g_x$ slightly.
The best estimates for the exact locations of the phase transition are $(\delta,K^c) = (0.6,1.327)$, $(1.0,2.0)$, $(1.4,2.664)$, which were determined by numerical extrapolations in the thermodynamic limit similar to inset in Fig.~\ref{fig:pd}.
}
\end{figure}

Through a somewhat different protocol than was used to find the phase diagram, we are able to estimate the central charge at the phase transition.
In this case we optimize MPS for the phase transition beginning from random initial states of small bond dimension; we then increase the bond dimension of the optimized state and re-converge, generating a finer series in bond dimension $\chi$.
As a result, individual data points are not independent of one another, although the data for differing $\delta$ are independent.

In Fig.~\ref{fig:central_charge_delta} we show results for the central charge measured at the phase transition along various cuts $\delta = 0.6,1.0,1.4$.
In this figure we have used the extrapolated critical values $K^c(\delta)$ and generated MPS for these points over a large range of bond dimensions $\chi$ from $30$ to $360$, corresponding to $\xi$ ranging from approximately $10$ to $200$.
We do not explicitly break any symmetries in this scheme, and individual data points within the figures for each value of $\delta$ are not independent.
The entanglement entropy measurements are consistent with the expected critical scaling $S = \frac c6 \ln \xi$, where $\xi$ is the correlation length induced in the wavefunction by the finite MPS bond dimension.

We find nearly the same central charge on the phase boundary at these points which are fairly widely separated.
This provides initial evidence that the phase boundary is controlled by a single fixed point, rather than a line of fixed points parameterized by a marginal operator as was found to be the case in Ref.~\cite{roberts2019deconfined}.
For values of $\delta$ close to 0 there is a crossover which interferes with the numerics, but otherwise this result is consistent with a single fixed point, reached by a flow from the decoupled $\Z_3$ criticality.

\subsection{Critical exponents} \label{sec:crit_exp}

With optimized MPS ground states in hand describing the phase transition, measuring correlation functions of lattice operators with suitable symmetry properties allows for the universality to be determined based on critical indices.
At a critical point various correlations display quasi-long-range order with asymptotic scaling $C_O(r) = \lrangle{O^\dag(0) O(r)} - \lrangle{O^\dag(0)} \lrangle{O(r)} \sim r^{-2 \Delta_O}$.

We will focus on the line $\delta=1$ and measure several correlations at the phase transition, including $Z_j$ which carries $g_x$ charge.
We also measure the \U1 current with temporal part $n_{1,j}$ and spatial part $j_{1,j}$ derived from the conservation of $\N_1$: explicitly, 
\begin{equation}
j_{1,j} \sim (-1)^j \left( T^1_j T^2_{j+1} + T^2_j T^1_{j+1} - T^6_j T^7_{j+1} - T^7_j T^6_{j+1} \right)~.
\end{equation}
In order to extract long-wavelength correlations of the conserved currents, we measure
\begin{equation}
C_{n_1}(r=j'-j) \equiv \lrangle{(n_{1,j} + n_{1,j+1}) (n_{1,j'} + n_{1,j'+1})}
\end{equation}
and similarly for $C_{j_1}(r)$.
We also measure $S^+_{1,j}$, which is charged under $\N_1$ but not $\N_2$:
\begin{equation}
S^+_{1,j} = \begin{bmatrix}0&p_j&0\\1-p_j&0&0\\0&0&0\end{bmatrix}.
\end{equation}
Again $p_j$ is the parity of $j$; see Eq.~\eqref{eq:p_j}.
The counterparts $n_{2,j}$, $j_{2,j}$, and $S^+_{2,j}$ are related to these operators by $\CC$.
These are all sensible for the transition at $\delta=1$; away from this line definite charge under $g_z$ is carried by $X_j$ or $X_j^\dag$, depending on $p_j$.
However $X_j$ and $X_j^\dag$ are simply linear combinations of the $\U1\times\U1$ raising and lowering operators as well as other terms related by permutation symmetry, which we expect is respected at the critical point.
So the critical exponent governing $S^+_{1,j}$ and $S^+_{2,j}$ will also determine the decay of correlations of $X_j$.
We confirmed the symmetry numerically but do not show these results, instead summarizing this family of operators by $S^+_{1,j}$ only, and similarly for $n_{1,j}$ and $j_{1,j}$.

We also measure the 0-momentum and $\pi$-momentum components of the energy term $E_j = \overline T^a_j T^a_{j+1}$ which is invariant under the full symmetry group:
\begin{align}
\epsilon^0_j &= E_j + E_{j+1}~, \\
\epsilon^\pi_j &= E_j - E_{j+1}~.
\end{align}
The operator $\epsilon^\pi_j$ is the natural lattice operator for \VBS correlations, being in the singlet sector of all internal symmetries (actually the entire \SU3) but odd under $\Z_2$ translation symmetry.

Finally, we wish to investigate the claim that the critical theory at the point $\delta=1$ in fact controls the entire phase boundary.
This would imply that the $\U1\times\U1$ symmetry of the line $\delta=1$ is emergent at the transition for other values of $\delta$; equivalently, terms breaking the symmetry are irrelevant at the transition for $\delta=1$.
We measure correlations of a term which carries charge under $\U1^2$ but preserves all symmetries of $H$ in Eq.~\eqref{eq:Hz3}.
None of the terms of the operator $\A = \sum_j A_j$ with the following $A_j$ preserve $\N_1$ or $\N_2$ quantum numbers, but $\A$ respects $g_z$, $g_x$, $\CC$, $\Theta$, and lattice symmetries:
\begin{equation}
A_j = \sum_{h \in S_3} \left(\ket{h(1)}\bra{h(0)}_j \otimes \ket{h(0)}\bra{h(2)}_{j+1} + \Hc \right).
\label{eq:A_op}
\end{equation}
The sum is over elements of the permutation group, and the term corresponding to the identity element $e = (012)$ is $S_{1,j}^+ S_{2,j+1}^+ + S_{1,j}^- S_{2,j+1}^-$.
We thus interpret $\mathcal A$ as driving $\U1\times\U1$ symmetry breaking while maintaining criticality.

Based on the above interpretation, we can predict the slope of the phase boundary in the phase diagram at $\delta=1$.
As mentioned there, the critical point $H^\ast$ appears to be located at $(\delta,K)=(1,2)$, where $J_z = K$.
Now we suppose that $\A$ turns out to be the most relevant symmetry-breaking operator, and moreover that $H^\ast + \lambda \A$ remains critical for small $\lambda$.
Decomposing this term into the $(\delta,K)$ basis, which control terms $(X_j X_{j+1} - Z^\dag_j Z_{j+1} + \Hc)$ and $6 \overline T^a_j T^a_{j+1}$, respectively, yields the unique solution
\begin{align}
A_j &= \left( X_j X_{j+1} + \frac 13 Z^\dag_j Z_{j+1} + \Hc\right) + 2 \overline T^a_j T^a_{j+1}\\
&= (X_j X_{j+1} - Z^\dag_j Z_{j+1} + \Hc) + \frac 53 \left(6 \overline T^a_j T^a_{j+1}\right)\nonumber \\
&\qquad\quad + \frac 43 \left( (Z^\dag_j Z_{j+1} + \Hc ) - 6 \overline T^a_j T^a_{j+1} \right)~. \label{eq:A_op_2}
\end{align}
The final line in Eq.~\eqref{eq:A_op_2} simply renormalizes $H^\ast$, allowing it to be removed from the perturbation term in this picture.
So as a consequence of the irrelevance of this $\U1^2$ symmetry-breaking term, we predict that the critical manifold in these variables has slope $\frac 35$ at $\delta=1$; this is highly consistent with the numerical data shown in Fig.~\ref{fig:pd}.

\subsubsection{Direct approach}

\begin{figure}[ht]
\includegraphics[width=\columnwidth]{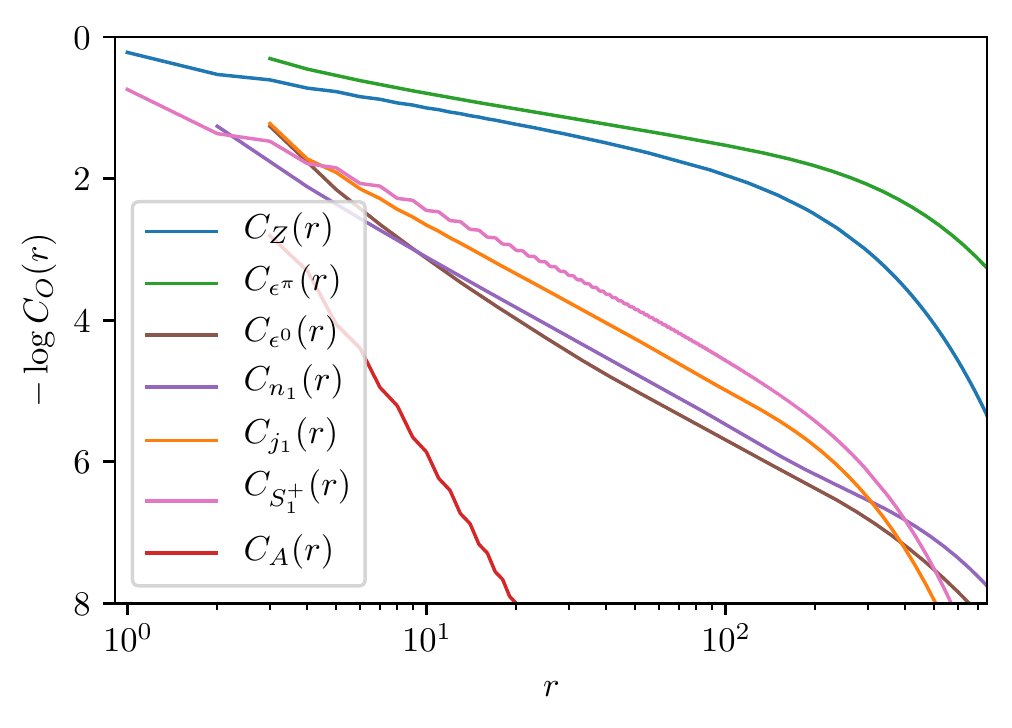}
\begin{tabular}{c||c|c|c|c|c|c|c}
$O$ & $Z$ & $\epsilon^\pi$ & $\epsilon^0$ & $n_1$ & $j_1$ & $S^+_1$ & $A$ \\
\hline
$~\Delta^d_O~$ & $~0.36~$ & $~0.37~$ & $~1.15~$ & $~1.17~$ & $~1.17~$ & $~0.89~$ & $~3.77~$
\end{tabular}
\caption{\label{fig:crit_exp_direct} 
Direct measurements of correlations are taken from an MPS of bond dimension $\chi=300$ optimized for the phase transition at $\delta = 1$, with translation invariance; that is, biased towards breaking $g_x$.
These operators are described in Sec.~\ref{sec:crit_exp}, and all correlations measure the connected component.
In the trace of $C_{\epsilon^0}$ we include only odd separations $r$ in the interest of visual clarity; the power law is unaffected. 
}
\end{figure}

The most straightforward approach to determining scaling dimensions is simply to measure the correlation function in real space and fit to a power law form.
We refer to this as the ``direct approach,'' following terminology used in Ref.~\cite{stojevic2015conformal}.
This is very similar to the procedure used in Ref.~\cite{roberts2019deconfined} to fit critical indices for the transition between Ising FM and \VBS; as was the case there, we determine a power law for the decay of correlations for a single bond dimension (usually the largest studied).
However, in contrast to that work we will always use the connected correlations; accordingly, we will not obtain bounds on exponents as we did there but rather simple estimates.
We suspect that this measurement will tend to overestimate operator scaling dimensions as a result of the finite length scale induced by the MPS bond dimension even at a critical point.
In addition, the direct approach suffers from ambiguity in determining the appropriate intermediate power-law region between non-universal short-distance behavior and eventual exponential decay.
We show the results of these measurements in Fig.~\ref{fig:crit_exp_direct}.

There is already an interesting observation visible in the raw data; namely, that the magnetic \zFM and \VBS observables have very similar power laws.
This is suggestive of some enhanced symmetry unifying the two order parameters at the putative critical point, a characteristic property of DQCP.

\subsubsection{Finite-entanglement scaling approach}

\begin{figure}[ht]
\includegraphics[width=\columnwidth]{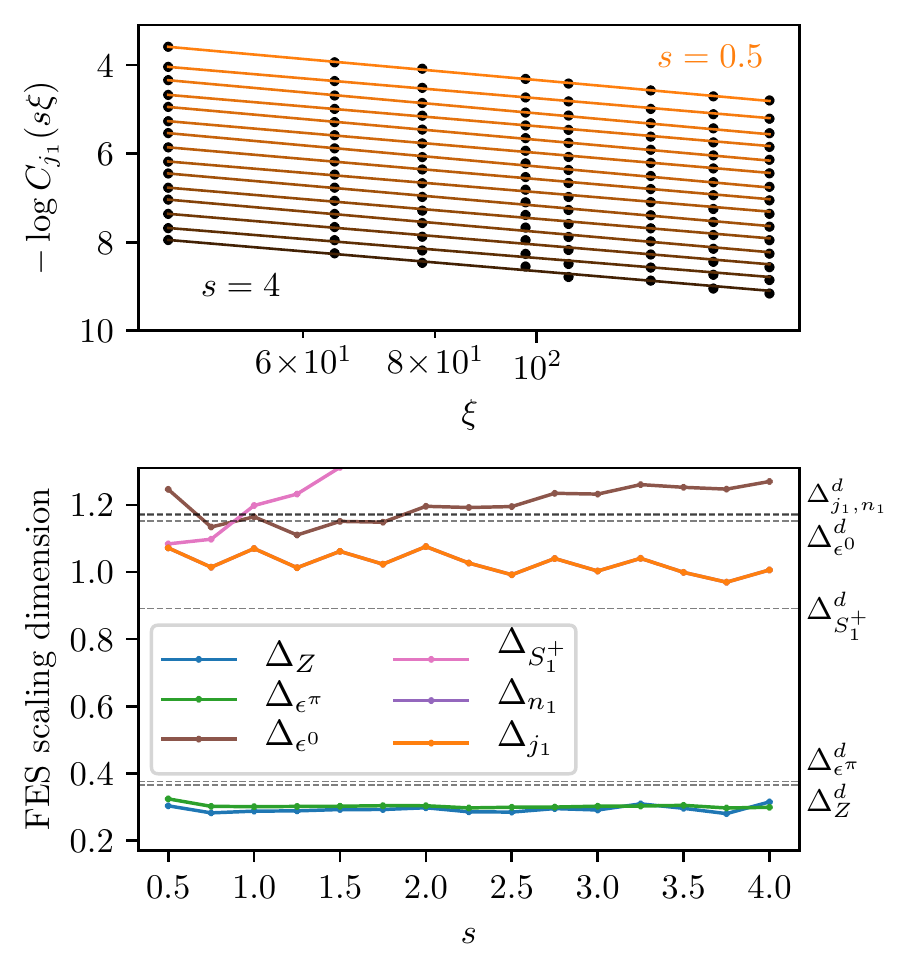}
\begin{tabular}{c||c|c|c|c|c|c}
$O$ & $Z$ & $\epsilon^\pi$ & $\epsilon^0$ & $n_1$ & $j_1$ & $S^+_1$ \\
\hline
$~\Delta_O(s=1)~$ & $~0.29~$ & $~0.30~$ & $~1.16~$ & $~1.07~$ & $~1.07~$ & $~1.20~$
\end{tabular}
\caption{\label{fig:crit_exp} In the FES approach we measure the correlations $C_O(s\xi)$ for a range of fixed dimensionless fractions $s$ and varying $\xi$.
The top panel shows data for the spatial part of the \U1 current $j_{1,j}$.
For $s > 1$ the raw data is already in the exponential decay regime of Fig.~\ref{fig:crit_exp_direct}, while this approach still exhibits consistent power law scaling; thus FES is indeed largely insensitive to the scaling function induced by finite MPS bond dimension.
In the bottom panel we show scaling dimensions as a function of $s$.
$\Delta_{{j_1}}$ and $\Delta_{{n_1}}$ are visually identical for all values of $s$.
We do not include $A$, which decays too quickly to use this method.
Horizontal lines marked $\Delta^d$ indicate values found by power-law fits in the direct approach in Fig.~\ref{fig:crit_exp_direct}.
In the table, we provide FES results at $s=1$.
}
\end{figure}

As mentioned previously, finite-entanglement approximations necessarily induce a length scale; here the MPS correlation length $\xi$ introduces some scaling function to the critical correlations which eventually decays exponentially. 
One specific technique to counteract this is referred to as ``finite-entanglement scaling'' (FES) \cite{stojevic2015conformal}, which is based on the observation that irrespective of the functional form of the correlations with a length scale, one finds that $C_O(s \xi) \sim (s \xi)^{-2\Delta_O}$.
Here $s$ is a dimensionless fraction which is kept fixed as one varies bond dimension (and hence $\xi$).
We employ this more sophisticated strategy which incorporates data from multiple optimized MPS in Fig.~\ref{fig:crit_exp}, and provide a comparison with the direct results.

One sees that the direct approach can tend to overestimate scaling dimensions as compared to FES, with the exception of the $S^+_{1,2}$ operators, whose raw data is not very amenable to a power-law fit.
Other results are qualitatively consistent with the direct approach, with highly relevant operators in the magnetic and translation symmetry--breaking sectors, along with other less-relevant operators charged under the \U1 symmetries and in the singlet sector.
The expectation that the conserved charges and currents $n_1$ and $j_1$ have scaling dimension 1 is reasonably well satisfied.
Additionally, the similarity between \zFM and \VBS order parameters is maintained in this approach, albeit with slower power laws.
The correlations $C_A$ decay too strongly to effectively treat with the FES method and are not shown.

From the scaling dimensions $\Delta_Z$, $\Delta_{\epsilon^\pi}$, and $\Delta_{\epsilon^0}$ measured in correlation functions we can provide numerical estimates of the critical indices characterizing the transition.
The FES scaling dimensions generally depend on $s$, and there is no {\it a priori} best value of this parameter to choose.
Fortunately our measurements do not vary widely, and for lack of a better option we will choose $s=1$.
These values are given in Fig.~\ref{fig:crit_exp}, and the reader is free to decide how seriously to take the numbers.
The order parameter exponents we compute are $\beta_\zFM \approx \beta_\VBS = 0.35$, and the correlation length exponent is $\nu = 1.2$.
Due to the strong irrelevance of the $\A$ perturbation breaking $\U1\times\U1$ symmetry, we predict that these critical indices describe an extended region of the phase boundary.

We revisit these measurements in Sec.~\ref{sec:ed} and compare with results from exact diagonalization, identifying these operators with primary fields in a putative CFT where possible.

\section{Theories of phase transition}
\label{sec:theories}

\subsection{Domain wall description} \label{sec:duality}

We write the standard duality mapping to $\Z_3$ domain wall variables on the dual lattice.
Denote the dual operators by $\ZZ_{j+1/2}$ and $\XX_{j+1/2}$:
\begin{align}
\XX_{j+1/2} &= Z^\dag_j Z_{j+1}~, \\
\ZZ_{j+1/2} &= \prod_{i\leq j} X_i~, \\
\ZZ^\dag_{j-1/2} \ZZ_{j+1/2} &= X_j~.
\end{align}
The dual operators satisfy $\ZZ \XX = \omega \XX \ZZ$.
In these variables $H$ is written (up to constant terms)
\begin{align}
\widetilde H &= - \sum_j \Big( (J^x \ZZ^\dag_{j-1/2} \ZZ_{j+3/2} + J^z \XX_{j+1/2} + \Hc) \nonumber\\
&\qquad\qquad\quad+ K (1 + \ZZ^\dag_{j-1/2} \ZZ_{j+3/2} + \Hc) \nonumber\\
&\qquad\qquad\qquad\qquad\quad~~\times(1 + \XX_{j+1/2} + \Hc)\Big)~,
\label{eq:Hdual}
\end{align}
and the generators of the $\Z_3^x \times \Z_3^z$ symmetry are expressed
\begin{equation}
g_x = \prod_j \ZZ^\dag_{j-1/2} \ZZ_{j+1/2} = 1~,~~g_z = \prod_k  \XX_{2k+1/2}~.
\end{equation}
That on a periodic chain $g_x$ appears trivial is a symptom of this duality failing to account for the global symmetry aspects of the model on such a chain.
In App.~\ref{app:z3_dual}, we formulate the duality on a periodic chain and account for all global aspects by using a dual $\Z_3$ gauge field.
We can view the analysis in this section as being performed in a fixed gauge.

The action of the symmetries on the dual variables is
\begin{align}
g_x:~~&(\XX_{j+1/2},\ZZ_{j+1/2})\mapsto(\XX_{j+1/2},\ZZ_{j+1/2})~,\\
g_z:~~&(\XX_{j+1/2},\ZZ_{j+1/2})\mapsto(\XX_{j+1/2},\omega^{p_j-1}\ZZ_{j+1/2})~,\\
\Theta:~~&(\XX_{j+1/2},\ZZ_{j+1/2})\mapsto(\XX^\dag_{j+1/2},\ZZ_{j+1/2}),~\ii \mapsto -\ii~,\\
\CC:~~&(\XX_{j+1/2},\ZZ_{j+1/2})\mapsto(\XX^\dag_{j+1/2},\ZZ^\dag_{j+1/2})~,\\
T_1:~~&(\XX_{j+1/2},\ZZ_{j+1/2})\mapsto(\XX_{j+3/2},\ZZ_{j+3/2})~,\\
\II:~~&(\XX_{j+1/2},\ZZ_{j+1/2})\mapsto(\XX_{-(j+1/2)},\ZZ_{-(j+1/2)})~.
\end{align}

The dual Hamiltonian Eq.~\eqref{eq:Hdual} can be viewed as two individual 3-state clock models residing on the ``even'' and ``odd'' sublattices of the dual lattice (locations $2k+1/2$ and $2k+3/2$, $k \in \mathbb{Z}$, respectively), with energy-energy coupling between them.
Physically, when all domain walls are gapped (that is, $\lrangle{\ZZ_\odd} = \lrangle{\ZZ_\even} = 0$) the \zFM order is preserved.
The threefold degeneracy of this phase is encoded in the gauge sector, which we have omitted in this writing but is presented in full in App.~\ref{app:z3_dual}.

Other phases can be obtained by various condensation patterns of the domain wall variables.
For example, condensing $\lrangle{\ZZ_\odd} = \lrangle{\ZZ_\even} \neq 0$ breaks $g_z$ but preserves $g_x$, $\CC$, $\Theta$, and $T_1$.
We thus identify this with the particular classical state $\bigotimes_j \ket{0_x}_j$ in the \xFM phase.
The other classical states in this phase break $\CC$ and $T_1$ but preserve $T_1 \CC$.
These correspond to $\lrangle{\ZZ_\odd} = \omega^{\pm 1} \lrangle{\ZZ_\even} \neq 0$.
It appears naively that there are a total of nine degenerate minima; however, when global symmetry aspects are accounted for by including the dual $\Z_3$ gauge field, there are indeed only three degenerate ground states.

By instead condensing domain walls as $\lrangle{\ZZ_\odd} \neq 0$ and $\lrangle{\ZZ_\even} = 0$, or vice versa, one finds a phase which breaks translation symmetry and has twofold ground state degeneracy.
We identify this condensate with the \VBS-ordered phase in the lattice model.
While this order parameter transforms nontrivially under $g_z$, its value is not gauge-invariant, and this phase indeed respects the full internal symmetry group.
From the perspective of the \zFM in this language, the \VBS phase is a Higgs phase, and the transition between these is given by condensing domain walls on only one sublattice of the dual lattice.

One can write a schematic theory of coarse-grained domain walls described by complex fields $w_A \sim \ZZ_\odd$, $w_B \sim \ZZ_\even$, transforming as
\begin{align}
g_x:~~&(w_A,w_B)\mapsto(w_A,w_B)~,\\
g_z:~~&(w_A,w_B)\mapsto(w_A,\omega^{-1} w_B)~,\\
\Theta:~~&(w_A,w_B)\mapsto(w_A,w_B),~\ii \mapsto -\ii~, \\
\CC:~~&(w_A,w_B)\mapsto(w_A^\dag,w_B^\dag)~,\\
T_1:~~&(w_A,w_B)\mapsto(w_B,w_A)~,\\
\II:~~&(w_A,w_B)\mapsto(w_A,w_B)~.
\end{align}
The associated Lagrangian reads
\begin{align}
\LL &= \LL_A + \LL_B + \LL_{AB}~, \\
\LL_\alpha &= t |w_\alpha|^2 + u_3 (w_\alpha^3 + c.c.) + u_4 |w_\alpha|^4 + \cdots~, \\
\LL_{AB} &= \lambda |w_A|^2 |w_B|^2 + \cdots~,
\label{eq:LAB}
\end{align}
where $\LL_\alpha$ is a schematic theory for the $\Z_3$ ordering transition on each sublattice.
Gradient terms are omitted for simplicity.
In addition to the usual mass term $t$ and quartic term $u_4$, the symmetries allow the $\Z_3$ anisotropy term $u_3$, which energetically distinguishes three particular directions to capture the qualitative physics of the underlying $\Z_3$ clock variables $\ZZ_{\odd/\even}$.

In the absence of coupling between the two sublattices, the critical point (on each sublattice) is obtained by tuning the parameter $t$.
Schematically, for ``renormalized'' $t_\text{renorm} > 0$ the fields $w_A$ and $w_B$ are both gapped, which for the original system corresponds to the \zFM phase.
In contrast, for $t_\text{renorm} < 0$ both fields condense; in the original system this corresponds to the \xFM phase.
This is not a tractable field theory for describing the $\Z_3$ criticality; instead, the actual critical properties are known from exact solutions of lattice models or study of the IR theory, which is a conformal minimal model.
Nevertheless, this schematic writing simplifies the discussion of the domain wall theory.

$\LL_{AB}$ represents coupling between the $\Z_3$ systems on the two sublattices.
In our model, this has the form of energy-energy coupling, for which we write the most relevant term with amplitude $\lambda$
\footnote{Additional terms in $\LL_{AB}$ in Eq.~\eqref{eq:LAB} can be obtained, {\it e.g.}, by forming symmetric combinations of products of terms in $\LL_A$ and $\LL_B$.
The listed symmetries allow terms like $\kappa[(w_A w_B)^3 + \Hc]$ and $\kappa'[(w_A^\dag w_B)^3 + \Hc]$ which individually are not energy-energy terms between the subsystems $A$ and $B$.
However, our specific lattice model in the dual formulation has an additional symmetry which acts like $\CC$ on one sublattice only; that is,
$\tilde{\CC}_A: \tilde{Z}_{2k-1/2} \mapsto \tilde{Z}_{2k-1/2}^\dagger, \tilde{Z}_{2k+1/2} \mapsto \tilde{Z}_{2k+1/2}, w_A \mapsto w_A^\dagger, w_B \mapsto w_B$.
This requires $\kappa=\kappa'$, and the combined term is an energy-energy term.
This minor difference between general models with the defined symmetries and our specific model is not used in any essential way.
The above additional symmetry of the lattice model which is manifest in the dual formulation is non-local in the original formulation.
}.
It is known from the CFT description of the $\Z_3$ criticality that the energy-energy coupling is relevant at the decoupled point.

Consider now the full theory including $\LL_{AB}$.
By lowering $t$, one allows domain walls to proliferate and destroy the \zFM order.
Focusing on the quartic terms, if $\lambda < 2u_4$ both domain walls want to condense simultaneously, leading to the \xFM phase.
(As described previously, the above Lagrangian does not include the dual $\Z_3$ gauge field needed to account for global symmetry aspects, which reduces to only three ground states.)
If instead $\lambda > 2u_4$ it is energetically favorable for only one domain wall species to condense, with two possibilities: either $\lrangle{w_A} \neq 0$, $\lrangle{w_B} = 0$ or $\lrangle{w_A} = 0$, $\lrangle{w_B} \neq 0$, which correspond to the two degenerate ground states of the VBS phase.

In our lattice model, the above two regimes correspond to $K < 0$, where we find a transition from the \zFM to the \xFM phase, and to $K > 0$, where we find the VBS phase.
Furthermore, along the $\delta=0$ line we find a first-order \zFM-\xFM phase boundary for $K<0$ while the VBS phase immediately opens up for $K>0$.
This is consistent with the relevance of the energy-energy coupling at the decoupled point $(\delta,K) = (0,0)$, taken together with the above schematic energetics picture of the preferred domain wall condensation patterns for $K < 0$ and $K > 0$.
Moreover, in our model along the line $\delta = 0$, the domain wall theory is invariant under a simultaneous duality transformation for each species $A$ and $B$, treated as their own $\Z_3$ chains, which we interpret as maintaining the ``thermal" variable $t_\text{eff} = 0$ and allowing only the energy-energy coupling to flow.
The runaway flows are then interpreted as leading to coexistence of \zFM and \xFM on one side---having $w_A$ and $w_B$ both gapped or both condensed being energetically equal by the above self-duality---and the VBS phase on the other side.

We can now discuss the \zFM-\VBS phase boundary, which requires perturbing from the decoupled point in both $t$ and $\lambda$ directions in the field theory (both $\delta$ and $K$ in our lattice model).
In the low-energy theory both couplings $t$ and $\lambda$ are relevant, with scaling dimensions $4/5$ and $8/5$, respectively.
The leading flow equations are $dt/d\ell = (6/5)t + \cdots$ and $d\lambda/d\ell = (2/5)\lambda + \cdots$ (in particular, $t(\ell) \sim \lambda(\ell)^3$ along the flows near the decoupled point).
To be on the phase boundary, the couplings $t$ and $\lambda$ must balance one another.
Thus we predict that the phase boundary has the shape $\delta_c(K) \sim K^3$ near the decoupled point.

Unfortunately, we do not know the ultimate fate of this type of balanced flow of two relevant couplings.
One possibility is that the flow leads to a new fixed point with only one relevant direction, which would then describe a generic continuous \zFM-\VBS transition.
The alternative is that there is no such new fixed point, and a runaway flow is interpreted as corresponding to a first-order \zFM-\VBS transition.
The above ``theory'' does not provide a controlled way to study this question, but we hope that it will motivate more interest in this problem.

\subsection{Theory for \texorpdfstring{$\U1\times\U1$}{U1xU1}-symmetric model}

\subsubsection{Bosonized variables}
The apparently emergent $\U1\times\U1$ symmetry invites treatment via bosonization.
This model can be approximated by two coupled $\U1$ rotors with variables $(n_{a,j},\phi_{a,j})$, $a=1,2$, defined by
\begin{align}
(-1)^j\ket{a}\bra{a}\sim n_{a,j}~,\quad
S_{a,j}^{+}
\sim \ee^{\ii\phi_{a,j}} ~,
\label{eq:luttinger_site_n_phi}
\end{align}
where $[n_{a,i},\phi_{a',j}]=\ii\delta_{aa'}\,\delta_{ij}$.

To begin writing the field theory description, we first determine the average filling in this system.
The filling number is constrained by $g_x$ action in Eq.~\eqref{eq:U1xU1_gx}; for a fully symmetric state we have
\begin{align}
\lrangle{n_{1,j}}=\lrangle{n_{2,j}}=\frac{(-1)^j}{3}~.
\label{}
\end{align}
Next, to capture fluctuations $\delta n_{a}\equiv n_a-\lrangle{n_a}$ we introduce bond variables $\theta_{a,j+1/2}$, where
\begin{align}
  \delta n_{a,j}=\frac{1}{\pi}\left( \theta_{a,j+1/2}-\theta_{a,j-1/2} \right)~.
  \label{}
\end{align}
We choose $\theta_{a,j+1/2}$ as follows:
\begin{align}
  \theta_{a,2k-1/2}&=\sum_{j'\le 2k-1}\pi\,n_{a,j'}~,\notag\\
  \theta_{a,2k+1/2}&=\sum_{j'\le 2k}\pi\,n_{a,j'}+\frac{\pi}{3}~.
  \label{eq:luttinger_bond_theta}
\end{align}
The commutator between $\theta_a$ and $\phi_{a'}$ is
\begin{equation}
[\theta_{a,j+1/2},\phi_{a',j'}]=\ii\pi\,\delta_{aa'}\,\Theta(j+1/2-j')~,
\end{equation}
where $\Theta(x)$ is the Heaviside step function.

To get to the low-energy theory, we define long-wavelength fields $\theta_{1,2}(x)$ and $\phi_{1,2}(x)$ in continuum space, where $\theta_{1,2}(x)$ are real-valued with periodicity $\pi$ and $\phi_{1,2}(x)$ have periodicity $2\pi$.
These fields satisfy
\begin{align}
  \left[ \frac{\partial_x\theta_a(x)}{\pi},\phi_{a'}(x') \right]=\ii\delta_{aa'}\,\delta(x-x')~.
  \label{}
\end{align}
The action of the symmetries on the fields can be deduced from their lattice counterparts in Eqs.~\eqref{eq:luttinger_site_n_phi} and \eqref{eq:luttinger_bond_theta}:
\begin{align}
  u(\varphi_1,\varphi_2)&: (\phi_1, \theta_1, \phi_2, \theta_2) \to (\phi_{1}+\varphi_{1}, \theta_1, \phi_2+\varphi_2, \theta_2)~,\nonumber\\
  g_x&:(\phi_1,\theta_1,\phi_2,\theta_2) \rightarrow (-\phi_1+\phi_2, \theta_2, -\phi_1, -\theta_1-\theta_2)~,\nonumber\\
  \Theta&:(\phi_1,\theta_1,\phi_2,\theta_2)\to (-\phi_1,\theta_1,-\phi_2,\theta_2)~,~~\ii\to-\ii~,\nonumber\\
  \CC&:(\phi_1,\theta_1,\phi_2,\theta_2)\to (\phi_2,\theta_2,\phi_1,\theta_1)~,\nonumber\\
  T_1&:(\phi_1,\theta_1,\phi_2,\theta_2)\to \left(  -\phi_1,-\theta_1+\frac{\pi}{3},-\phi_2,-\theta_2+\frac{\pi}{3} \right)~,\nonumber\\
  \II&:\left( \phi_1(x),\theta_1(x),\phi_2(x),\theta_2(x) \right)\to\nonumber\\
  &\left(\phi_1(-x),-\theta_1(-x)+\frac{\pi}{3},\phi_2(-x),-\theta_2(-x)+\frac{\pi}{3} \right)~.
\label{eq:luttinger_sym}
\end{align}

We are now ready to write down the low-energy theory.
The Gaussian part reads
\begin{align}
  \LL_0=&\sum_{a=1}^2 \left[ \frac{\ii}{\pi}\partial_\tau\phi_a\partial_x\theta_a+\frac{v}{2\pi}\left( g(\partial_x\phi_a)^2+\frac{1}{g}(\partial_x\theta_a)^2 \right) \right]\notag\\
  &+\frac{v}{2\pi}\left( -g\partial_x\phi_1\partial_x\phi_2+\frac{1}{g}\partial_x\theta_1\partial_x\theta_2 \right),
  \label{eq:luttinger_gaussian}
\end{align}
with a single tunable Luttinger parameter $g$ and velocity $v$.
There are two types of symmetric scattering terms: 
\begin{widetext}
  \begin{enumerate}
    \item Type I:
      \begin{align}
        \lambda^I_m\left[ \cos\left(2m(\theta_1+\theta_2)-\frac{2m\pi}{3} \right) +\cos\left( 2m\theta_1+\frac{2m\pi}{3} \right)+\cos\left( 2m\theta_2+\frac{2m\pi}{3} \right) \right]~,\quad
        m\in \Z ~;
        \label{eq:luttinger_sym_scatter_I}
      \end{align}
    \item Type II:
      \begin{align}
        \lambda^{II}_m\left[ \cos\left( 2m(\theta_1-\theta_2) \right)+\cos\left( 2m(\theta_1+2\theta_2) \right)+\cos\left( 2m(2\theta_1+\theta_2) \right) \right]~,\quad
        m\in \Z ~.
        \label{eq:luttinger_sym_scatter_II}
      \end{align}
  \end{enumerate}
\end{widetext}

The scaling dimensions for generic exponentials of the fields at the Gaussian fixed point are given by \cite{KaneFisher1995}:
\begin{align}
  &\dim\left[ \exp(\ii (2m_1\theta_1+2m_2\theta_2 )) \right]=\frac{2g}{\sqrt{3}}(m_1^2-m_1m_2+m_2^2)~,\notag\\
  &\dim\left[ \exp(\ii(p_1\phi_1+p_2\phi_2)) \right]=\frac{1}{2\sqrt{3}g}(p_1^2+p_1p_2+p_2^2)~.
  \label{}
\end{align}
We now specialize the above, listing some important operators in this bosonized language along with scaling dimensions at the Gaussian fixed point.
\begin{itemize}
  \item As discussed before, operators carrying unit charge under $\U1\times\U1$ are $S_{1,2}^+ \sim \exp(\ii\phi_{1,2})$, which have scaling dimensions $\dim[S_{1,2}^+]=\frac{1}{2\sqrt 3 g}$.
  \item The operator $A$ defined in Eq.~\eqref{eq:A_op}, which breaks $\U1\times\U1$ to $\Z_3^z$, reads
    \begin{align}
      A\sim \cos(\phi_1 + \phi_2) + \cos(2\phi_1 - \phi_2) + \cos(\phi_1 - 2\phi_2) ~,
      \label{eq:luttinger_A_term}
    \end{align}
    and $\dim[A]=\frac{\sqrt{3}}{2g}$.
  \item The \zFM order parameter is given by
    \begin{align}
      O_\zFM\sim&\cos\left(2\theta_1+2\theta_2-\frac{2\pi}{3} \right) + \ee^{2\ii\pi/3}\cos\left( 2\theta_1+\frac{2\pi}{3} \right)\notag\\
      &\qquad+\ee^{-2\ii\pi/3}\cos\left( 2\theta_2+\frac{2\pi}{3} \right)~,
      \label{}
    \end{align}
    and $\dim[O_\zFM]=\frac{2g}{\sqrt 3}$ at the Gaussian fixed point.
  \item The \VBS order parameter reads
    \begin{align}
      O_\VBS\sim&\cos\left(2\theta_1+2\theta_2-\frac{\pi}{6} \right) + \cos\left( 2\theta_1+\frac{\pi}{6} \right)\notag\\
      &\qquad+\cos\left( 2\theta_2+\frac{\pi}{6} \right)~,
      \label{}
    \end{align}
    and $\dim[O_\VBS]=\frac{2g}{\sqrt 3}$ at the Gaussian fixed point.
\end{itemize}

It is interesting to note that at the Gaussian fixed point, the \zFM and \VBS order parameters have the same scaling dimension, which also coincides with the scaling dimension of the leading allowed scattering term (type I in Eq.~\eqref{eq:luttinger_sym_scatter_I} with $m=1$).
Furthermore, we have the relation
\begin{equation}
\frac{\dim[S_a^+]}{\dim[A]} = \dim[S_a^+]\dim[\zFM] = \frac 13~.
\label{eq:gaussian_scaling_dims}
\end{equation}

When $g > \sqrt{3}$, all allowed scattering terms are irrelevant and this system is in a stable gapless phase described by the Gaussian fixed point, with power law exponents as described above.
This phase is stable as long as the $\U1\times\U1$ symmetry is present microscopically.
(On the other hand, if the $\U1\times\U1$ symmetry is broken down to $\Z_3^z$ and the $A$ term is allowed, one cannot simultaneously make this term and all scattering terms irrelevant and the gapless phase is unstable.)

\subsubsection{Gapped phases and classical phase diagram}
We now develop the representation of various gapped phases in this theory.
Different gapped quantum phases correspond to different patterns of $\lrangle{\phi_{1,2}}$ or $\lrangle{\theta_{1,2}}$.
As a consequence of the Mermin-Wagner theorem, $\phi_{1,2}$ never condense and we always have $\lrangle{\exp(\ii\phi_{1})}=\lrangle{\exp(\ii\phi_{2})}=0$.

For quantum states preserving $T_1$, we require $\lrangle{\theta_{1,2}}=\pi/6$ or $-\pi/3\pmod \pi$.
For quantum states preserving $g_x$, we require $\lrangle{\theta_1}=\lrangle{\theta_2}=0$ or $\pm\pi/3\pmod \pi$.
We are then able to represent the gapped phases appearing in the previous sections as follows:
\begin{itemize}
\item $\lrangle{\theta_1}=\lrangle{\theta_2}=-\pi/3$ gives a fully symmetric phase. 
The detailed study of this phase is presented in App.~\ref{app:spt_U1xU1}.
\item $\lrangle{\theta_1}=\lrangle{\theta_2}=0$ or $\pi/3$ gives the two degenerate ground states of the \VBS phase.
\item $(\lrangle{\theta_1},\lrangle{\theta_2})\!=\!(\pi/6,\pi/6),(\pi/6,-\pi/3),(-\pi/3,\pi/6)$ gives the three degenerate \zFM ground states.
\end{itemize}

The classical phase diagram of this Luttinger liquid theory is obtained by minimizing the energy of the scattering terms. 
We first consider the symmetric scattering term in Eq.~\eqref{eq:luttinger_sym_scatter_I} with $m=1$:
\begin{align}
\lambda_1^I\Bigg[ & \cos\left(2(\theta_1+\theta_2)-\frac{2\pi}{3} \right) + \cos\left( 2\theta_1+\frac{2\pi}{3} \right) \nonumber \\
&\qquad\qquad\quad+\cos\left( 2\theta_2+\frac{2\pi}{3} \right) \Bigg] ~.
\label{eq:luttinger_1st_sym_scattering}
\end{align}
Its scaling dimension is $2g/\sqrt{3}$, the lowest among symmetric terms; it is relevant for $g<\sqrt{3}$.
When $\lambda_1^I < 0$, Eq.~\eqref{eq:luttinger_1st_sym_scattering} is minimized at $\theta_1=\theta_2=-\pi/3$, and thus gives the symmetric phase.
When $\lambda_1^I > 0$, Eq.~\eqref{eq:luttinger_1st_sym_scattering} is instead minimized at $\theta_1=\theta_2=0$ or $\pi/3$, and thus gives the \VBS phase.
The Gaussian part of the two-component Luttinger liquid theory in Eq.~\eqref{eq:luttinger_gaussian} describes a possible phase transition from the \VBS phase to the symmetric phase.
More specifically, if we also have $g > 1/\sqrt{3}$ so that the next scattering term--- Eq.~\eqref{eq:luttinger_sym_scatter_II} with $m=1$---is irrelevant, the SPT to \VBS transition is obtained when the single relevant coupling $\lambda_1^I$ changes sign and is indeed described by the Gaussian theory.
The correlation length exponent at this transition is set by the scaling dimension of the $\lambda_1^I$ term: $\nu = 1/(2 - 2g/\sqrt{3})$, while the power law correlations of various observables are governed by the scaling dimensions we have calculated.
(It is interesting that even though \zFM order is not present on either side of the transition, its correlations decay with the same power law as the \VBS order present on one side.)

To describe the \zFM phase and its transition to the \VBS phase, we add the next scattering term to Eq.~\eqref{eq:luttinger_1st_sym_scattering}:
\begin{widetext}
\begin{align}
\lambda_1^I \Big[ \cos\left(2\theta_1+2\theta_2-\frac{2\pi}{3} \right) + \cos\left( 2\theta_1+\frac{2\pi}{3} \right) +&\, \cos\left( 2\theta_2+\frac{2\pi}{3} \right) \Big] \nonumber\\
&~~+\lambda_1^{II} \Big[ \cos\left(2\theta_1-2\theta_2 \right) + \cos\left( 2\theta_1+4\theta_2 \right) + \cos\left( 4\theta_1+2\theta_2 \right) \Big]~.
\label{eq:luttinger_two_cos}
\end{align}
\end{widetext}
When $g<1/\sqrt{3}$, both terms are relevant.

\begin{figure}[ht]
\includegraphics[width=\columnwidth]{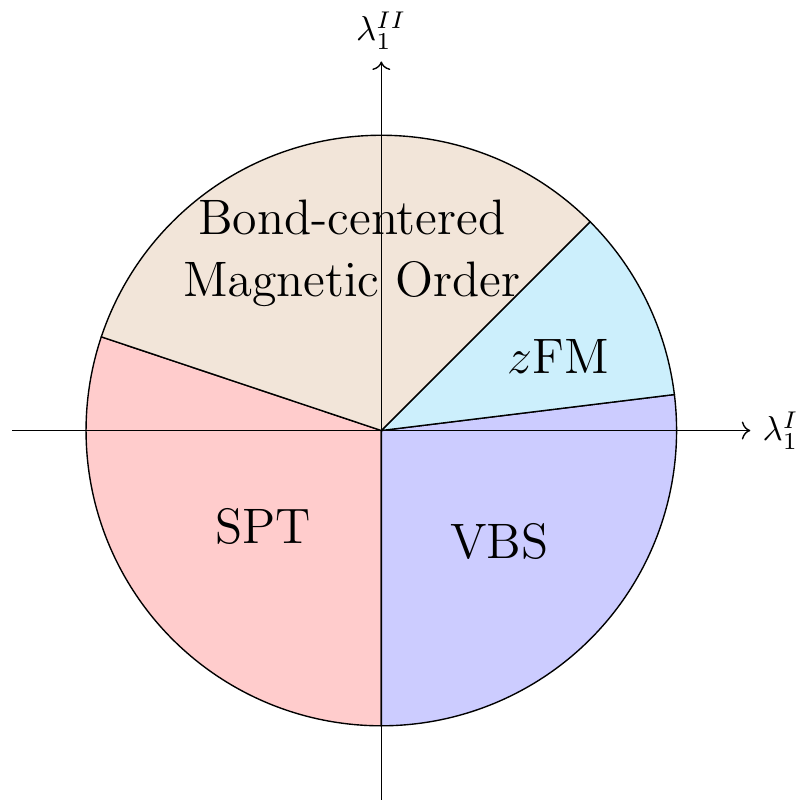}
\caption{\label{fig:luttinger_classical_phase_diagram} Four distinct phases appear in the classical phase diagram obtained by analyzing the minima of Eq.~\eqref{eq:luttinger_two_cos}.
}
\end{figure}

We parameterize $\lambda_{1}^{I,II}$ by $\lambda$ and $\alpha$, where $\lambda_1^I=\lambda\cos\alpha$ and $\lambda_1^{II}=\lambda\sin\alpha$.
For each $\alpha$, we identify all minima of Eq.~\eqref{eq:luttinger_two_cos}, and associate classical phases with the minima by analysis of symmetry properties.
The resulting phase diagram is shown in Fig.~\ref{fig:luttinger_classical_phase_diagram}.

When $\arctan(1/8)<\alpha\leq\pi/4$, then $(\theta_1,\theta_2)_{\min}=(\pi/6,\pi/6)$, $(\pi/6,-\pi/3)$ or $(-\pi/3,\pi/6)$, which gives the \zFM phase.
We are also able to identify representative lattice wavefunctions for these three states by studying their transformation properties under $\CC$ and $g_x$:
\begin{align}
&\left(\frac{\pi}{6},\frac{\pi}{6}\right) \sim \bigotimes_j \ket{0}_j~,~~\left(\frac{\pi}{6},-\frac{\pi}{3} \right) \sim \bigotimes_j\ket{1}_j~, \nonumber\\
&\left( -\frac{\pi}{3},\frac{\pi}{6} \right)\sim \bigotimes_j\ket{2}_j~.
\label{eq:Lutt2zFM}
\end{align}

When $-\pi/2<\alpha<\arctan(1/8)$, we find $(\theta_1,\theta_2)_{\min}=(0,0)$ or $(\pi/3,\pi/3)$, which gives the \VBS phase.

When $-\pi-\arctan(1/3)\leq\alpha<-\pi/2$, $(\theta_1,\theta_2)_{\min}=(-\pi/3,-\pi/3)$, and we find the symmetric phase.

When $\pi/4<\alpha<\pi-\arctan(1/3)$, we get six degenerate minima, which can be parameterized by a single variable $\upsilon$:
\begin{align}
(\theta_1,\theta_2)_{\min} = &\left( \frac{\pi}{6}\pm\upsilon,\frac{\pi}{6}\mp\upsilon \right),~
\left(\frac{\pi}{6}\mp\upsilon,-\frac{\pi}{3}\right), \nonumber\\
&\left(-\frac{\pi}{3},\frac{\pi}{6}\pm\upsilon\right).
\label{eq:bond_order_phase}
\end{align}
The physical picture of this phase can be obtained by analyzing the symmetries of these minima and their relation to nearby phases.
Denoting the above minima as $\mathfrak{A}_\pm, \mathfrak{B}_\pm, \mathfrak{C}_\pm$, they transform in a 3-cycle way under $g_x: \mathfrak{A}_\pm \to \mathfrak{B}_\pm \to \mathfrak{C}_\pm \to \mathfrak{A}_\pm$, while they are exchanged pairwise under lattice translation $T_1$ and inversion about a site $\mathcal{I}: \mathfrak{A}_+ \leftrightarrow \mathfrak{A}_-, \mathfrak{B}_+ \leftrightarrow \mathfrak{B}_-, \mathfrak{C}_+ \leftrightarrow \mathfrak{C}_-$.
Furthermore, $\mathfrak{A}_\pm/\mathfrak{B}_\pm/\mathfrak{C}_\pm$ are exchanged pairwise under symmetries $\CC$, $g_x \CC$, or $\CC g_x$.
At the point $\alpha=\pi/4$, the optimal $\upsilon=0$ and these pairs merge to give the three ground states of the \zFM phase in Eq.~\eqref{eq:Lutt2zFM}.
We conclude that the phase with $\upsilon \neq 0$ also has magnetic order similar to \zFM with additional translation and site inversion symmetry breaking (but preserves bond inversion symmetry).
However, the lattice symmetry breaking is different from the \VBS order: {\it e.g.}, the \VBS order parameter is zero in all these states for any $\upsilon$, and, more directly, the \VBS ground states are invariant under $\CC$  and $g_x$, which is not the case here.
According to the symmetry properties of this phase, we name it a ``bond-centered magnetic order'' phase.

We cannot write simple product states that would have the desired transformation properties, including the expected quantum numbers under the $\U1^2$.
However, it is possible to write MPS wavefunctions for these ground states, by building upon the MPS wavefunction for the neighboring SPT phase from App.~\ref{app:spt_U1xU1}, with which the present phase connects at $\alpha=\pi-\arctan{(1/3)}$, $\upsilon=\pi/2$, where all of the minima collapse to $(-\pi/3,-\pi/3)$ (remember that the $\theta$ fields are defined modulo $\pi$).
The MPS construction for this phase is presented in App.~\ref{app:phase_connect_zFM_spt}.

\subsubsection{\texorpdfstring{\zFM-\VBS}{zFM-VBS} transition in \texorpdfstring{$\U1\times\U1$}{U(1)xU(1)}-symmetric theory}
We can now discuss the phase transition between the \zFM and \VBS phases within this theory.
In the above ``classical" treatment of the scattering terms $\lambda_1^I$ and $\lambda_1^{II}$, the phase transition occurs along the line $\lambda_1^{II} = \lambda_1^I/8$ with positive $\lambda_1^{I,II}$; this is a ``level crossing" transition and is first order.
This treatment is appropriate when both bare couplings $\lambda_1^I$ and $\lambda_1^{II}$ are large.
On the other hand, we can consider starting from the Gaussian theory when these bare couplings are small. In the regime $g < 1/\sqrt{3}$, both couplings are relevant and start flowing to larger values.
We may speculate that the (almost) continuous \zFM to \VBS transition observed in our numerical study occurs when these couplings during their flow balance each other in just the right way, but unfortunately we do not have controlled means to study this.

Nevertheless, it is intriguing that some of the relations among the various scaling dimensions at the Gaussian fixed point appear to be approximately satisfied in our numerical study at the (pseudo-)critical point $(\delta, K) = (1, 2)$.
Namely, we find numerically that the \zFM and \VBS order parameters have very close scaling dimensions, while they are equal in the Gaussian theory.
We also find that Gaussian theory relations in Eq.~\eqref{eq:gaussian_scaling_dims} are approximately satisfied.
The scaling dimensions are consistent with a naive estimate $g_\text{eff} \approx 0.25$.
For such $g_\text{eff}$, both $\lambda_1^I$ and $\lambda_1^{II}$ would be relevant (in fact, one more scattering term $\lambda_2^I$ would also be relevant), consistent with these couplings flowing away from the Gaussian fixed point.
For such a value of $g_\text{eff}$, the term $A$ breaking the $\U1\times\U1$ symmetry down to $g_z$ is irrelevant, which is consistent with the observed emergent $\U1\times\U1$ symmetry along the \zFM-\VBS phase boundary.

We remark that the above relations among various exponents in the Gaussian theory follow from the fact that there is a single Luttinger parameter in the theory, which in turn is dictated by the microscopic symmetries.
It is possible that the corresponding approximate relations found in the numerical study of the (pseudo-)critical point are also primarily due to the symmetries rather than proximity to the specific two-component Luttinger liquid theory.
However, we do not know how to guess a better description, while the Luttinger liquid theory at least provides some framework for discussing observables and noticing these relations.

\section{Connection to integrable statistical mechanics models} \label{sec:integrable}

\subsection{Classical model of non-intersecting strings} \label{sec:nis}

Focusing on the line of enhanced symmetry $\delta=1$ which has significantly informed our study so far, one observes in Fig.~\ref{fig:pd} that this slice appears to intersect the phase boundary exactly at the point $(\delta,K)=(1,2)$, at which $J^x = 0$ and $J^z = K$.
Up to constants and an overall scale, this point is equivalent to
\begin{equation}
H^\ast = -\sum_j \Big((q-2)\!\sum_\alpha \ket{\alpha \alpha}\! \bra{\alpha\alpha}_{j,j+1} + \sum_{\alpha,\beta}\ket{\alpha \alpha}\! \bra{\beta \beta}_{j,j+1}\Big),
\label{eq:Hstar}
\end{equation}
for $q=3$.
This Hamiltonian may be special, and in order to understand it we first return to another special instance of our Hamiltonian, namely, the point $J^x = J^z = 0$, which up to normalization and constants maps exactly to the pure biquadratic spin-1 Hamiltonian $H_\mathrm{bQ}$, Eq.~\eqref{eq:Hbq}.
This Hamiltonian is also associated with the transfer operator of a particular two-dimensional statistical mechanics model realizing ``non-intersecting strings'' (NIS).

\begin{figure}[ht]
\setlength{\unitlength}{1cm}
\begin{picture}(10,3.5)
  \put(0,0.5){\includegraphics[width=\columnwidth]{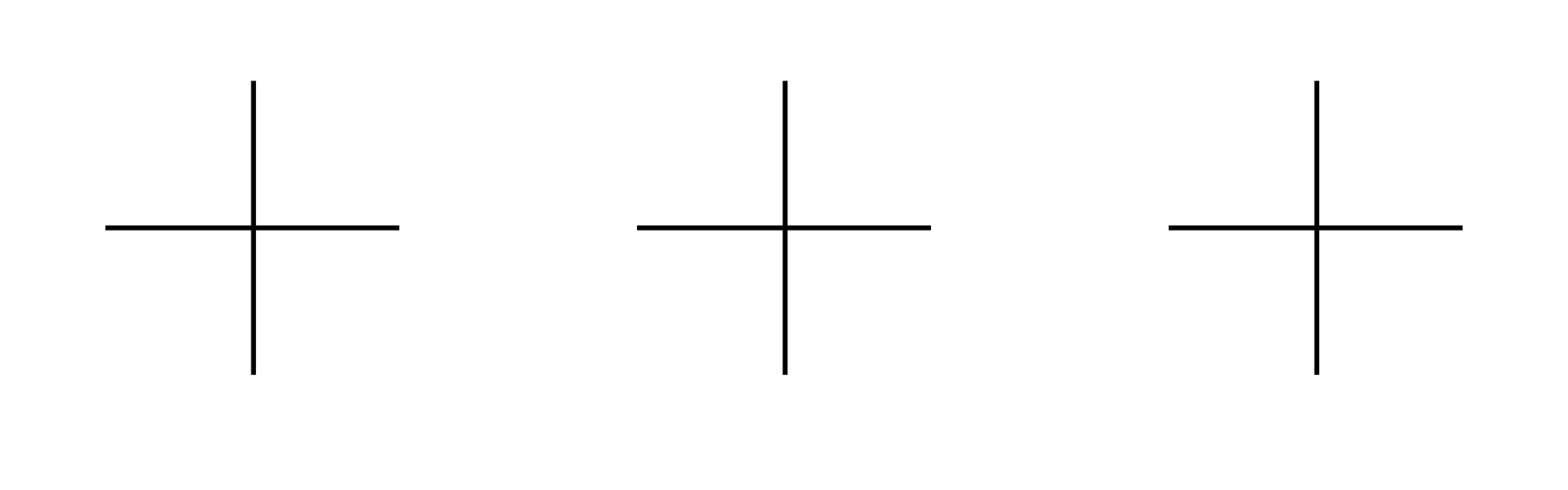}}
  \put(1.27,0.78){\normalsize $\alpha$}
  \put(0.22,1.88){\normalsize $\alpha$}
  \put(1.27,2.95){\normalsize $\alpha$}
  \put(2.35,1.88){\normalsize $\alpha$}
  \put(1.12,0.1){\large $(a)$}
  \put(4.22,0.78){\normalsize $\alpha$}
  \put(3.17,1.85){\normalsize $\beta$}
  \put(4.22,2.95){\normalsize $\beta$}
  \put(5.3,1.88){\normalsize $\alpha$}
  \put(4.08,0.1){\large $(c)$}
  \put(7.17,0.78){\normalsize $\alpha$}
  \put(6.12,1.88){\normalsize $\alpha$}
  \put(7.17,2.95){\normalsize $\beta$}
  \put(8.25,1.85){\normalsize $\beta$}
  \put(7,0.1){\large $(d)$}
\end{picture}
\caption{\label{fig:nis_vertex} The three types of vertices shown here, with $\alpha \neq \beta$, are allowed in the vertex models we consider.
We consider the model on the two-dimensional square lattice with vertex weights $a$, $c$, and $d$ for the configurations $(a)$, $(c)$, and $(d)$ respectively; see text for details.
}
\end{figure}
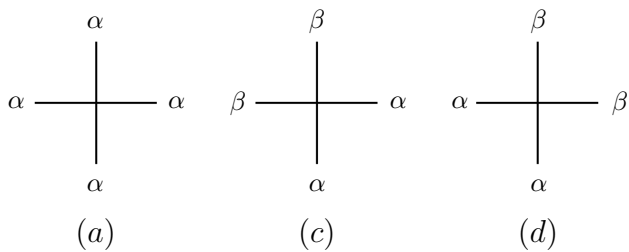

These models can be formulated with classical $q$-state degrees of freedom assigned to the edges of a graph---we will have in mind the two-dimensional square lattice---and weights assigned to the vertices according to their configurations.
The only nonzero vertices are those shown in Fig.~\ref{fig:nis_vertex}; when accounting for the $S_q$ permutation symmetry of the labels $\alpha,\beta = 1,\ldots,q$, there are $q(2q - 1)$ allowed vertices.
To simplify the notation, we write the weights as $w_{(a)} = a$, $w_{(c)} = c$, and $w_{(d)} = d$ \footnote{In choosing these vertex labels and weights we follow the convention of \citet{klumper1990spectra}.}.
Solving the Yang--Baxter equation for the transfer matrix with $S_q$ symmetry yields two integrable models for each value of $q$, satisfying the following conditions \cite{schultz1981solvable,perk1981new,perk1986graphical}:
\begin{align}
\text{separable:}\quad 
& a = c + d~,\label{eq:sep_cond}\\
\text{non-separable:}\quad
& a^2 = a(c+d) + (q-2)cd~. \label{eq:ns_cond}
\end{align}
The solution Eq.~\eqref{eq:sep_cond} is commonly known as the separable NIS model, and we refer to that of Eq.~\eqref{eq:ns_cond} as the integrable non-separable case.

Schematically, under the separability condition Eq.~\eqref{eq:sep_cond}, vertices of type $(a)$ can be decomposed into both types $(c)$ and $(d)$ and thereby removed from the partition sum.
Then one can map via a two-step duality to the self-dual point of the $q^2$-state Potts model \cite{perk1986nonintersecting}.
The $q^2$-state Potts degrees of freedom reside on alternating plaquettes of the original square lattice and have generally anisotropic nearest-neighbor interactions in the $\hat x + \hat y$ and $\hat x - \hat y$ directions of the NIS lattice, with Boltzmann weights set by $c/d$ and $d/c$.
For any $c$ and $d$ the model is self-dual; the point $c = d$ corresponds to the isotropic self-dual model.
We provide the explicit duality mapping from the separable $q$-state NIS model on the square lattice to the $q^2$-state Potts model, as well as further discussion, in App.~\ref{app:potts_duality}.

The quantum Hamiltonian associated with the $q=3$ separable model is $H_\mathrm{bQ} \propto H[J^x=0,J^z=0,K]$ in the phase diagram of Eq.~\eqref{eq:Hz3}.
As discussed in App.~\ref{app:su3}, this model is known to be among the relatively few integrable spin-1 models and can be understood through either a duality mapping to a 9-state self-dual Potts model or by appeal to equivalent Temperley--Lieb models.

In the integrable, but non-separable, NIS model the $(a)$ vertex cannot be removed, and the operator algebra of the associated quantum Hamiltonian includes a corresponding non-Temperley--Lieb generator.
As a result, we are not aware of any useful algebraic equivalences to well-known models which could expose the low-energy properties of this model.
This quantum Hamiltonian associated with the integrable non-separable NIS model for $q=3$ is in fact $H^\ast$, Eq.~\eqref{eq:Hstar}.

Both integrable NIS statistical mechanics models are exactly solvable for general $q$ by the analytic Bethe ansatz \cite{perk1983diagonalization,perk1986graphical}.
The structure is quite similar to the solution of the XXZ model using magnons, with the reference states of the method being the highest excited states (a manifold spanned by $\ket{\alpha_1,\alpha_2,\dots,\alpha_N}$ with $\alpha_i \neq \alpha_{i+1}$).
Although the solution for the eigenvalues was performed explicitly by \citet{de1993exact}, it is not known how to access the low-energy subspace or ground state wavefunctions exactly.

\subsection{Phases of NIS models}

The weight of a single vertex can be written (with link variables labeled in the compass pattern S,W,N,E)
\begin{align}
w(&\alpha,\gamma,\beta,\rho) \nonumber \\
&= a\,\delta_{\alpha\gamma\beta\rho} + c\, (\delta_{\alpha\rho}\delta_{\beta\gamma} - \delta_{\alpha\gamma\beta\rho}) + d\, (\delta_{\alpha\gamma}\delta_{\beta\rho} - \delta_{\alpha\gamma\beta\rho}) \nonumber\\
&= (a - c - d)\, \delta_{\alpha\gamma\beta\rho} + c\, \delta_{\alpha\rho}\delta_{\beta\gamma} + d\, \delta_{\alpha\gamma}\delta_{\beta\rho}~.
\label{eq:nis_vertex}
\end{align}
Since the overall scale of $w$ does not change the probabilities, the vertex model has two independent parameters, which we are free to choose.
We use $c/d$, which characterizes lattice anisotropy, as well as another parameter characterizing the relative weight of the $(a)$-type vertices compared to the $(c)$- and $(d)$-type vertices.
One choice for such a parameter would be $a^2/cd$, but we will instead use a related quantity,
\begin{equation}
\Theta = \frac{a}{cd}(a - c - d) = \frac{a^2}{cd} - \frac{a}{\sqrt{cd}} \left(\sqrt{\frac{c}{d}} + \sqrt{\frac{d}{c}} \right)~.
\label{eq:theta}
\end{equation}
The parameter $\Theta$ is convenient in that the two integrable models correspond to $\Theta=0$ and $\Theta=q-2$.
At each of these special values of $\Theta$, the NIS transfer matrices commute for any anisotropy parameter $c/d$; this is simply a restatement of the Yang-Baxter solubility of these models. 
In particular, the information encoded in the eigenvectors of the transfer matrices is independent of the ``spectral variable'' $c/d$.
Accordingly, we can say that the physics is strictly independent of the anisotropy parameter.
This conclusion does not hold at other values of $\Theta \neq 0, q-2$ and the quantitative details will depend on the anisotropy; however, we expect that the qualitative physics will still be independent.

Using the freedom afforded by the spectral variable, one can tune to the extreme anisotropic limit of the $\Theta = 0, q-2$ transfer matrices and take a logarithmic derivative to determine that these integrable models yield precisely the $H_\text{bQ}$ and $H^\ast$ quantum Hamiltonians, respectively \cite{schultz1981solvable,perk1981new,perk1983diagonalization,klumper1989new,klumper1990spectra}.
In this section we will allow $\Theta$ to vary and will argue that $\Theta < q-2$ realizes the same phase as the separable model $\Theta=0$ which breaks the lattice translation symmetry, while $\Theta > q-2$ realizes a magnetically ordered phase.
Hence, the integrable non-separable model $\Theta = q-2$ appears to be at the transition between these phases.

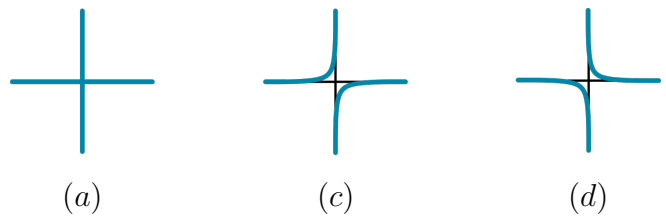
\begin{figure}[ht]
\setlength{\unitlength}{\columnwidth}
\begin{picture}(1,0.4)
  \put(0,0.1){\includegraphics[width=\columnwidth]{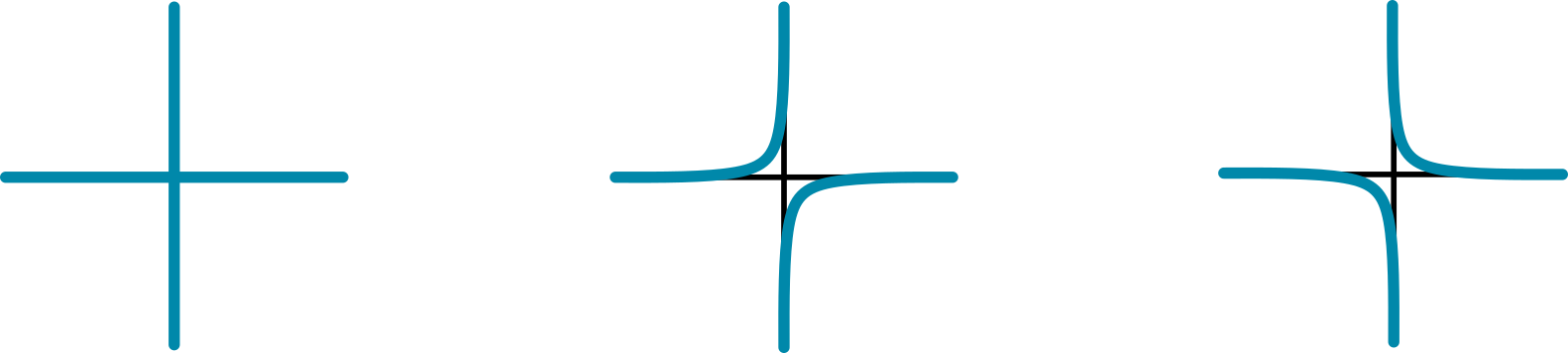}}
  \put(0.08,0.02){\large $(a)$}
  \put(0.472,0.02){\large $(c)$}
  \put(0.858,0.02){\large $(d)$}
\end{picture}
\caption{\label{fig:nis_vertex_loop} The vertex configurations of the loop model, which are unflavored, are shown.
The weight of a configuration depends only on the geometric pattern of connections of the string segments assigned to the edges of the two-dimensional square lattice.
The weight of each individual vertex type can be read off from Eq.~\eqref{eq:nis_vertex}; the partition sum in terms of such loops is specified in Eq.~\eqref{eq:Zloop}.
}
\end{figure}

As suggested by its name, the NIS model partition sum can be rewritten in terms of nonlocal strings; these are ``completely packed'' on the square lattice, with each edge containing a string segment.
Every vertex can connect the segments on its adjoining edges in three different ways according to the pictures of $(a)$-, $(c)$-, and $(d)$-type vertices in Fig.~\ref{fig:nis_vertex_loop}.
Ignoring boundaries, one sees that allowed string configurations take the form of loops lying along connected edges, all of which are in the same state within a single loop.
These loops may self-intersect at $(a)$-type vertices but do not cross one another.
The partition function can then be rewritten independently of the $q$ possibilities for the state of the edges comprising each loop, and the sum over flavors performed explicitly, obtaining a model in which $q$ appears as a parameter and weights in the partition sum are determined entirely by loop geometry.
The precise formulation in terms of unflavored strings is akin to a high-temperature expansion for a $q$-state Potts model.
The utility of this formulation is that treating $q$ as a parameter specifying a loop fugacity allows it to be varied continuously.

The weights of these vertices can be read off from Eq.~\eqref{eq:nis_vertex}, so by substituting for $\Theta$ using Eq.~\eqref{eq:theta} we write the general partition function in terms of the loops:
\begin{align}
Z &= \sum_\sigma q^{\ell(\sigma)} (a-c-d)^{n_a(\sigma)} c^{n_c(\sigma)} d^{n_d(\sigma)} \label{eq:Zloop} \\ 
&= (cd)^{\frac{N}{2}} \sum_\sigma q^{\ell(\sigma)} \left(\!\sqrt{\Theta + \gamma^2} - \gamma \right)^{n_a(\sigma)}\!\left(\frac{c}{d}\right)^{\frac{n_c(\sigma)-n_d(\sigma)}{2}}, \nonumber
\end{align}
where $\gamma$ is determined from the anisotropy by
\begin{equation}
\gamma \equiv \frac{1}{2} \left(\sqrt{\frac{c}{d}} + \sqrt{\frac{d}{c}} \right) \geq 1 ~.
\end{equation}
(The isotropic point with $\gamma = c/d = 1$ is a one-parameter loop model.)
In the partition sum $\sigma$ denotes a configuration of completely packed unflavored loops with connections drawn from Fig.~\ref{fig:nis_vertex_loop} at the vertices.
Here $\ell(\sigma)$ is a nonlocal quantity, namely the number of loops in $\sigma$, and $n_a$, $n_c$, and $n_d$ are the numbers of vertices of each type in $\sigma$.
The NIS model defined on the oriented lattice coincides with the model defined on the unoriented lattice for $c=d$; thus, the results about integrability still hold along this line.
However the staggered model with $c \neq d$ does not have commuting transfer matrices even for $\Theta=0,q-2$.

Consider first a regime in which the $(a)$ vertex is suppressed at low energies.
Setting $\Theta = 0$ enforces $n_a(\sigma)=0$ identically.
As mentioned earlier, this model is equivalent to the $q^2$-state Potts model, with anisotropic couplings if $c \neq d$, but such that self-duality is maintained.
For $c=d$, the model is isotropic and for $q>2$ is known to be at a first-order transition between the Potts ordered and disordered phases (and we expect this to be true also for $c \neq d$).
In the NIS language, the ordered and disordered phases of the Potts model are known to correspond to short-loop states running predominantly around one or the other set of plaquettes \cite{perk1986nonintersecting,perk1986graphical,affleck1990exact,Note100}.
\footnotetext[100]{The most direct way to see that the ordered and disordered phases of the Potts model correspond to one or the other checkerboard pattern of NIS loops is to consider the isotropic NIS model and perturb it by staggered weights for the $(c)$ and $(d)$ vertices, oppositely for the two sublattices.
In the NIS language this selects one of the checkerboard states, while under the duality to the $q^2$-state Potts model this moves the Potts model off self-duality and hence into one of the phases.}
This is a ``checkerboard'' phase of the loop model which spontaneously breaks the lattice symmetry, but is symmetric under $S_q$ permutation of the labels.
The general model has $\Theta \neq 0$, allowing $(a)$ vertices.
Presumably the short-loop checkerboard phase is stable under introducing some finite amount of $\Theta$.
(In the language of the related $q^2$-state Potts model with $q^2 > 4$, a small $\Theta$ perturbation moves along a first-order coexistence line.)
This is the VBS phase of our spin model.

Conversely, in a regime with high weight on the $(a)$ vertex, configurations at low energies include strings that extend across the whole system.
In the language of the original vertex model degrees of freedom, such proliferation of strings corresponds to spontaneous breaking of the $S_q$ permutation symmetry by choosing one of the $q$ colors.
Thus, the phase will display long-range correlations of a magnetic-type order parameter which measures whether distant links are connected by an unbroken string, whereas in the short-loop checkerboard phase correlations of this order parameter decay exponentially.
In our spin model, the proliferated-loop phase is the \zFM phase.

Now for intermediate values of the parameter $\Theta$ there will be a transition between the extended phase and both short-loop phases.
Our finding that the \VBS to \zFM transition in the $q=3$ model appears to be exactly at the integrable point corresponding to $\Theta=q-2$ suggests that the completely packed loop model we describe undergoes a transition between checkerboard short loops and the proliferated loop phase at exactly $\Theta = q-2$.
A similar conjecture was made in Ref.~\cite{wang2015completely} in the context of special completely packed O($n$) loop models (which map precisely onto the above loop model with $q=n$) and is supported by transfer matrix studies for $n \geq 10$ and $n < 2$.
As we discuss in the next subsection, the $\Theta=q-2$ model actually has a finite correlation length, which however can be enormous for $q \gtrsim 2$, of which our spin model with $q=3$ is an example. 
Our DMRG study reaching correlation lengths around 200 and locating the \zFM-\VBS transition very close to the point $\Theta=q-2$ gives very strong support to this conjecture also in the vicinity of $q=3$.

\subsection{Walking description of phase transition}

\subsubsection{Summary of exact results for integrable models}

There is a way to learn about the spectrum of the transfer matrix of the integrable NIS models without the need to construct eigenstates, through the so-called inversion trick introduced by \citet{stroganov1979new} and later used to study the six-vertex model by \citet{baxter1982inversion,baxter1982exactly}.
In its initial setting the inversion relation was actually developed specifically to compute the free energy per site of the two integrable $q=3$ NIS models, before more was known about their structure.
An extended inversion relation was used by \citet{klumper1989new,klumper1990spectra} to compute subleading eigenvalues of the transfer matrix, exposing some details of the low-energy spectrum.
In particular, he found that the dependence on $q$ of the thermodynamic-limit energy gaps of both quantum Hamiltonians corresponding to the integrable NIS models (under some overall normalization) is governed by the function
\begin{equation}
\Delta = g(x) = \log x \; \prod_{n=1}^\infty \left(\frac{1 - x^{-n/2}}{1 + x^{-n/2}}\right)^2~,
\end{equation}
and the correlation length by $\xi = f(x)$ \cite{klumper1988eigenvalues,klumper1990spectra},
\begin{equation}
f(x) = -1/\log k(x)~,~~k(x) = \frac{4}{\sqrt x} \prod_{n=1}^\infty \left( \frac{1+x^{-2n}}{1+x^{-2n+1}}\right)^4.
\label{eq:k_exact}
\end{equation}
The two integrable models correspond to the following functional forms of the argument $x$:
\begin{align}
x_\sep(q) &= \frac{q + \sqrt{q^2-4}}{q - \sqrt{q^2-4}}~, \\
x_\ns(q) &= q-1~.
\end{align}

One can draw some conclusions about these models from the equivalence between the separable $q$-state NIS model and the $q^2$-state self-dual Potts model.
Because the self-dual Potts model transitions from critical to gapped at $Q_\Potts = 4$, then $\Delta^\sep = 0$ for $q \leq 2$ and $\Delta^\sep > 0$ for $q > 2$.
Thus we can also determine the value $q^c$ at which $\Delta^\ns$ experiences a transition from gapless to gapped.
Because $x_\sep(q=2) = 1 \equiv q^c - 1$, in fact the non-separable NIS model also experiences a transition from gapless to gapped at the value $q^c=2$.
In particular, using $q=3$ and the normalization from Sec.~\ref{sec:pd}, we exactly determine the energy gap of the Hamiltonian $H^\ast$ to be $\Delta = 1.42 \times 10^{-4}$ and the correlation length $\xi = 190878$ lattice spacings.
From the point of view of the functions $g(x)$ and $f(x)$, this is because the integrable non-separable lattice model has the gap and correlation length which correspond to the self-dual Potts model with $Q_\Potts = [x_\sep^{-1}(x_\ns(q=3))]^2 = \frac 92$.
The $Q_\Potts = 5$ model is known to already have a large correlation length of 3553 lattice spacings, and $Q_\Potts = \frac 92$ is even closer to the critical value $Q_\Potts^c = 4$.

To recapitulate the content of this section, the $q$-state separable integrable NIS model maps to the self-dual Potts model with $Q_\Potts = q^2$ states, and this mapping is actually an equivalence of models in the bulk (that is, ignoring boundary effects).
On the other hand, in the $q$-state non-separable integrable NIS model, the expression for the gap and correlation length are those which also apply to a Potts model at $Q_\Potts = [x_\sep^{-1}(x_\ns(q))]^2 = q^2/(q - 1)$, but we could not find any arguments for a stronger equivalence between these models.

\subsubsection{Implications for renormalization group flow}

Supposing that the $q=3$ non-separable NIS model indeed describes the phase boundary, one concludes that the transition is {\it extremely} weakly first-order.
The emergence of such a length scale enormously greater than the lattice spacing presents a ``hierarchy problem.''
Fortunately we can again look to the self-dual Potts model which provides a more familiar example of this phenomenon.
In the preceding section we used exact results for the eigenvalues of the transfer matrix to contextualize the very small gap and long correlation length of $H^\ast$ in terms of the Potts pseudo-criticality.
Much is now known about the Potts case due to a recent thorough treatment as an instance of ``walking'' of renormalization group flows \cite{gorbenko2018walking,gorbenko2018walking2}.

In brief, walking is the following proposal of an RG equation for a microscopic coupling $\lambda$:
\begin{equation}
\frac{d \lambda}{d \log L} = -\epsilon + \lambda^2 + \cdots~.
\label{eq:rg_flow}
\end{equation}
For $\epsilon > 0$ the flow has fixed points $\lambda^\ast = \pm \sqrt \epsilon$, one of which is stable and the other unstable.
(In the Potts case these are the critical and tricritical points existing at  $Q_\Potts < 4$; the system is assumed to be already tuned to the phase transition, {\it e.g.}, by enforcing the self-duality, and $\lambda$ is some remaining parameter in this manifold.)
These fixed points merge upon tuning $\epsilon \to 0$, and ``disappear'' for $\epsilon < 0$.
However in this regime solutions $\lambda^\ast = \pm i \sqrt{|\epsilon|}$ still exist, and represent a particular type of non-unitary theory.
Quantities like central charge, scaling dimensions, and OPE coefficients at these complex fixed points generally have nonzero imaginary components, and the conformal data of the two fixed points are related by complex conjugation.

While the complex fixed points are inaccessible to RG flows in the unitary theory, they do control the physics at intermediate length scales.
This is because the running of the coupling slows down considerably near $\lambda = 0$ \footnote{One can treat solutions $\lambda^\ast$ with finite real part by simply removing it via a shift to $\lambda-\mathrm{Re}[\lambda^\ast]$.}, where it passes close to these ``complex CFTs.''
The RG time required for $\lambda$ to flow from $-1$ to $+1$ is found by integrating Eq.~\eqref{eq:rg_flow}: the result is $t \sim \frac{\pi}{\sqrt{|\epsilon|}}$, corresponding to a length scale \cite{gorbenko2018walking}
\begin{equation}
\xi = \xi_0 \exp \frac{\pi}{\sqrt\epsilon}~. \label{eq:walking_xi}
\end{equation}
For small values $|\epsilon| \ll 1$ this scale already becomes very long; in this case the approximate conformal symmetry inherited from the complex CFTs looks nearly exact even for large finite systems.
However, because the flow is not approaching a conformally symmetric fixed point, the conformal data measured in systems with a characteristic length scale will drift with the scale, displaying the eventual limiting behavior at a scale comparable to $\xi$.

In the self-dual Potts model the form of Eq.~\eqref{eq:rg_flow} is well motivated by a long history of study, with parameter 
$\epsilon_\Potts = \frac{1}{\pi^2} (4 - Q_\Potts)$ to leading order in the limit $Q_\Potts \to 4$ \cite{gorbenko2018walking}.
By matching the characteristic walking behavior at $\epsilon = 0$ with the divergent parts of the exact results in the previous section we can write down $\epsilon$ also for the non-separable model.
The function $k$ defined in Eq.~\eqref{eq:k_exact}, an elliptic modulus, can equivalently be written $k(x) = (\vartheta_2 (\q)/\vartheta_3 (\q))^2$, where $\vartheta_n(\q)$ is the Jacobi theta function $\vartheta_n(z=0,\q=1/x)$.
We emphasize that the usage of the letter $\q=1/x$ in this way is an unfortunate coincidence arising from the conventions of elliptic functions.

To leading order as $\q \nearrow 1$ (that is, from the weakly first-order side), we expand
\begin{equation}
\frac{\vartheta_2 (\q)}{\vartheta_3 (\q)} \approx 1 - \frac{4}{2 + \exp \left[ \frac{\pi^2}{1-\q}\right]}~,
\end{equation}
so $\log f(x) \sim \frac{\pi^2}{1-\q}$, and consequently
\begin{align}
\log f(x_\text{sep}(q)) &\sim \frac{\pi^2}{2\sqrt{q-2}} ~,\\
\log f(x_\text{ns}(q)) &\sim \frac{\pi^2}{q-2} ~,
\end{align}
to leading order in the limit $q \to 2$.
We therefore propose that in the RG equation for the integrable NIS models $\epsilon$ has the form
\begin{align}
\epsilon_\sep &= -\frac{4}{\pi^2}(q-2) ~, \label{eq:epsilon_sep}\\
\epsilon_\ns &= -\frac{1}{\pi^2}(q-2)^2~,~q \geq 2~. \label{eq:epsilon_ns}
\end{align}
These statements are strictly applicable only in the limit $q \to 2$ \footnote{Specifically, the correspondences between the integrable $q=3$ NIS models and the Potts models at $Q_\Potts=9$ and $\frac 92$ is not evident here due to the approximation.}.
In this limit, Eq.~\eqref{eq:epsilon_sep} reproduces the known result for the self-dual Potts model with $Q_\Potts = q^2 \approx 4 + 4(q-2)$; in particular, the complex fixed points separate as the square root of the deviation from the critical value of $q$: $\lambda^\ast_\text{ns} = \frac 2\pi \sqrt{q-2}$.
On the other hand, Eq.~\eqref{eq:epsilon_ns} indicates that the functional dependence on $q$ is different in the non-separable case: the next correction to $\log f(q-1)$ is a constant, so $\frac{d\epsilon}{dq} = 0$ at $q=2$ and $\lambda^\ast = \pm \frac i\pi (q-2)$ grows linearly with $q$.
By taking these results seriously at $q=3$---which is dubious based on the expansion but works well for the Potts model nonetheless; see Sec.~3.5 of Ref.~\cite{gorbenko2018walking}---from Eq.~\eqref{eq:walking_xi} one arrives at a value $\xi_0 \approx 9.9$ for $H^\ast$, which can be compared with the UV length scale $\xi_{0,\Potts} \sim 0.19$ obtained for the weakly first-order Potts transition.

In order to follow the standard story of walking $\epsilon_\ns$ should change sign at $q=2$; it may indeed be the case that, for instance, an additional factor of $\text{sign}(q-2)$ is required in Eq.~\eqref{eq:epsilon_ns}.
However, we observe that close to the marginal value $q=2$ the two separable and non-separable stories of walking we have been telling independently actually merge.
In our spin model the former case lies inside the VBS phase with fairly large correlation length $\xi \approx 21$ for $q=3$, diverging for $q \to 2$, while the latter resides on the \VBS-\zFM boundary and has a much larger correlation length with stronger divergence as $q \to 2$.
It is interesting that both of these points occur in the same NIS model as $\Theta$ is varied, and it is intriguing to speculate that the walking parameter $\lambda$ posited separately for each case may in fact be the same.
If this is true, the complex CFTs discussed for the two models occur in the same larger parameter space which also contains the parameter $\Theta$, and in principle a richer flow structure involving these fixed points is possible.
It would be interesting to address this speculation with more concrete calculations and also to examine possible implications for crossovers in the physical spin problem.

\section{Exact diagonalization study of CFT data of the integrable model} \label{sec:ed}

In the walking picture the physics of our model in the approximately conformal regime is dictated by complex CFTs; accordingly, numerics are well suited to illuminate some of the properties of these theories.
In order to do so we will study the lattice model using exact diagonalization (ED), where the details of the low-energy spectrum under periodic boundary conditions provides a reliable way to identify CFT operators, up to finite-size corrections \cite{milsted2017extraction}.
Specifically,
\begin{equation}
E_\alpha = \frac{2\pi}{N a}\left( \Delta_\alpha - \frac{c}{12} \right) + O(N^{-x})~,~~P_\alpha = \frac{2\pi}{N a} S_\alpha~,
\end{equation}
under suitable normalization of the lattice Hamiltonian.
(The lattice spacing is denoted $a$.)
Here $x > 1$ is a non-universal exponent controlling the finite-size scaling.
In this way we can also compare ED data with some of the results of Sec.~\ref{sec:vumps} by identifying the low-energy excitations associated with primary operators in the CFT.
The application of this idea to lattice models was first worked out by Koo and Saleur \cite{koo1993representations} for Bethe-ansatz integrable models and later developed into a more general numerical technique \cite{milsted2017extraction}.

\begin{figure}[ht]
\includegraphics[width=\columnwidth]{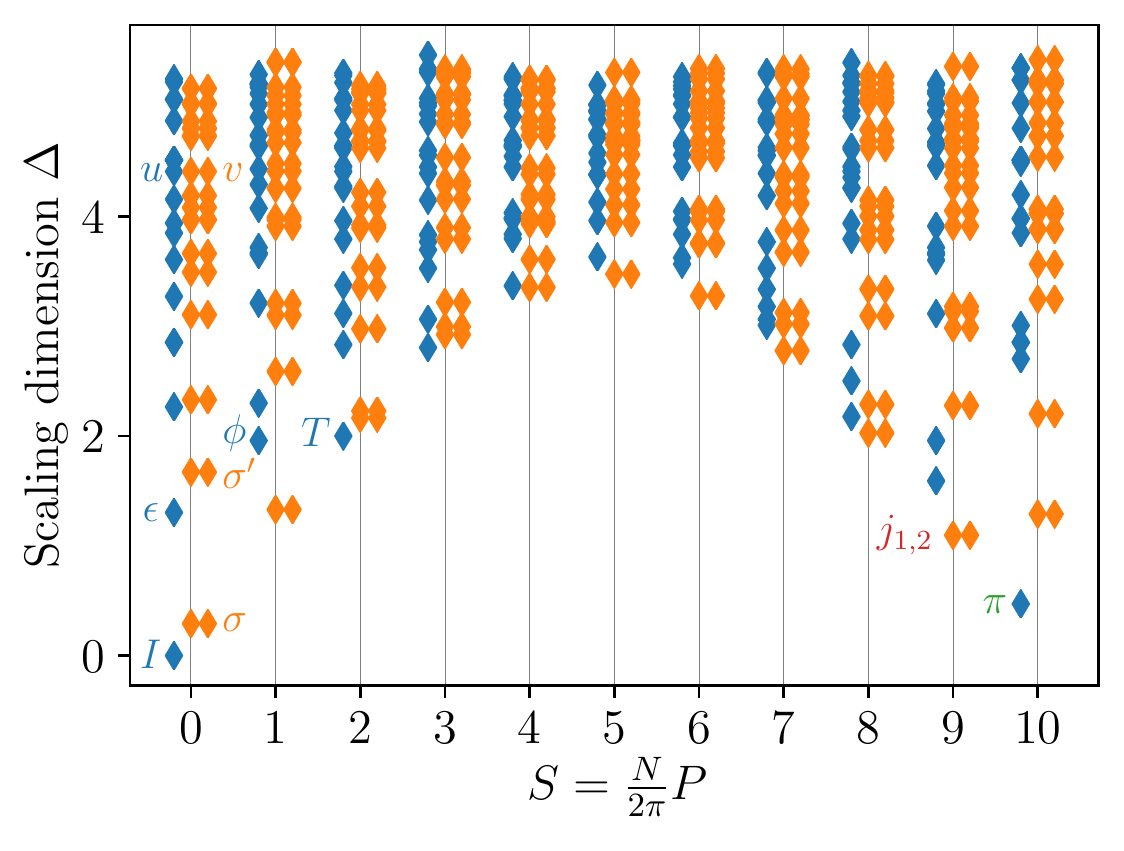}
\caption{\label{fig:ed_spec} We show the low-energy spectrum of the integrable model with system size $N=20$ in the $\N_1 = \N_2 = 0$ sector.
Eigenvalues are organized based on conformal spin $S$ and $g_x$ quantum number, with $g_x=1$ shown in blue and $g_x=\omega,\omega^2$ (which are related by $\CC$) in orange.
States are offset slightly from their quantized momenta for visual clarity.
Scaling dimensions $\Delta$ are determined by normalization of the energy eigenvalue of the $\ket{T}$ state associated with the stress-energy tensor, as $\Delta_T = 2$.
Highest-weight states identified using Fourier modes $H_n$ are indicated by name.
Quantum numbers of these states under other symmetries $\CC$ and $\II$ (where applicable) are not shown here but are listed in Table \ref{tab:primaries}.}
\end{figure}

The fundamental idea is based on the observation that the Fourier modes of the Hamiltonian density in a CFT on a circle are linear combinations of the Virasoro generators:
\begin{equation}
H_n^\text{CFT} = \frac{N a}{2\pi}\!\int_0^{N a}\!dx\,e^{i n x \frac{2\pi}{N a}} h^\text{CFT}(x) = L_n + \overline L_{-n},~n \neq 0.
\label{eq:fourierH}
\end{equation}
The action of a Virasoro (anti)chiral operator $L_n$ ($\overline L_n$) is to decrease (increase) conformal spin by $n$ and decrease conformal dimension by $n$. 
That is, $H_n^\text{CFT}$ imparts conformal spin $-n$, connecting lattice momentum sectors $\frac{2\pi}{N a}S$ and $\frac{2\pi}{N a}(S-n)$.
It is an elementary property of conformally symmetric theories that all states are grouped into conformal towers related by the Virasoso generators.
Each tower descends from a unique highest-weight state, which is associated with a primary field by the state-operator correspondence.
Because the energy of a state in the theory on a circle depends on the operator scaling dimension, the highest-weight states can be identified with those whose overlap with lower-energy states upon application of $H_n$ vanishes or goes to 0 with increasing $N$.
The numerical method obtains by applying these statements about continuum fields to the lattice operators, in particular assuming that the relationship Eq.~\eqref{eq:fourierH} also applies to Fourier modes of the lattice Hamiltonian and lattice counterparts of the Virasoro generators, up to finite-size corrections.

\begin{figure}[ht]
\includegraphics[width=\columnwidth]{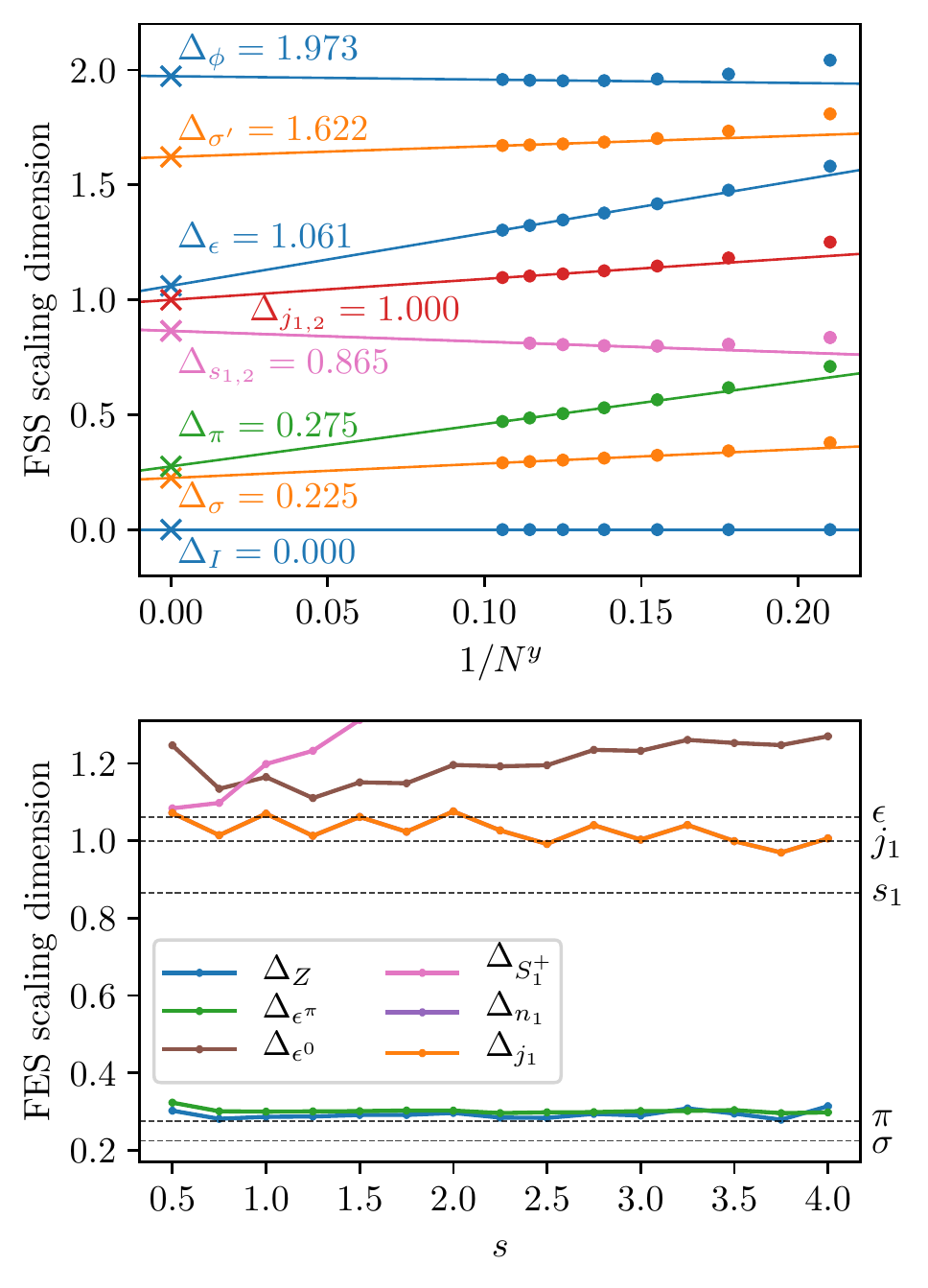}
\caption{\label{fig:fes_ed_exp} In the upper panel we show scaling dimensions of primary fields in the putative conformal fixed points obtained using finite-size scaling of the excitation energies of highest-weight states.
We determine the exponent $y = 3/4$ numerically, by observation of finite-size corrections to the vanishing matrix elements of $H_n$ with the state $\ket T$ used for normalization.
We do not show the relatively heavy operators $u,v$, but these behave similarly.
For the fits we use only system sizes $N \geq 12$, though also show data for $N=8,10$.
In the lower panel we repeat the plot containing data for the critical exponents obtained from the FES method, also shown in Fig.~\ref{fig:crit_exp}.
Now the horizontal lines marked on the figure indicate the scaling dimension of the most relevant primary field in each associated symmetry sector as measured in ED.}
\end{figure}

Based on the above, one does not need to construct lattice equivalents of the Virasoro generators; simply acting repeatedly with $H_n$, $n \in \{-2,-1,1,2\}$, on an eigenstate generates other states in the same conformal tower.
By projecting the lattice Fourier modes $H_n$ into the space of low-energy eigenstates, the structure of the conformal towers can be easily read off from the matrix elements, and those having zero matrix element for all $H_n$ with all eigenstates of lower energy must be the highest-weight states associated with primary fields in the CFT.
We find in our data that for some eigenstates this sum of matrix elements on lower-energy states vanishes identically.
In other cases an eigenstate may have a small matrix element which decreases with system size; if the spectrum does not contain another state from which this state could reasonably descend, we also label this state a primary and attribute the nonzero values of $H_n$ to finite-size corrections.
However, we are generally conservative and are not trying to exhaustively label all highest-weight states in the spectrum, but rather identify those that correspond to measurements made in previous sections, in addition to other obvious candidates.

\begin{figure}[ht]
\includegraphics[width=\columnwidth]{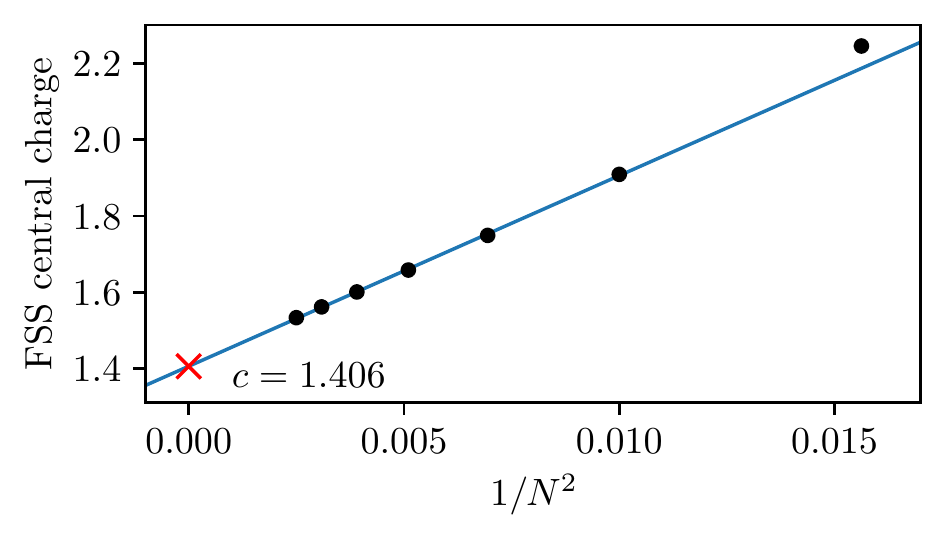}
\caption{\label{fig:fss_c} Finite-size scaling for the central charge is based on the matrix element $\bra T H_{-2} \ket I$, where $\ket I$ is the ground state and $\ket T$ the state with conformal spin $S=2$ associated with the stress-energy tensor in the field theory.
This state has the lowest energy in its sector for all system sizes studied.
The scaling with $N^{-2}$ is used for other models \cite{milsted2017extraction}, and visually appears to be appropriate.
The fit excludes the first two data points $N=8,10$.}
\end{figure}

\begin{table}[ht]
\begin{tabular}{c||c|c|c|c|c|c}
Primary field & $\text{Re}[\Delta]$ & $S$ & $\U1^2$ & $g_x$ & $\CC$ & $\II$\\
\hline
$I$ & $0$ & $0$ & $0$ & $0$ & $+$ & $+$\\
$\sigma,\tilde\sigma$ & $0.225$ & $0$ & $0$ & $\pm 1$ & & $+$\\
$\pi$ & $0.275$ & $N/2$ & $0$ & $0$ & $+$ & $-$\\
$s_1,s_2$ & $0.865$ & $N/2$ & $1_1,1_2$ & & & $+$ \\
$j_1,j_2$ & $1.000$ & $N/2 - 1$ & $0$ & & & \\
$\epsilon$ & $1.061$ & $0$ & $0$ & $0$ & $+$ & $+$\\
\hline
$\sigma',\tilde\sigma'$ & $1.622$ & $0$ & $0$ & $\pm 1$ & & $+$\\
$\phi,\overline \phi $ & $1.973$ & $\pm 1$ & $0$ & $0$ & $-$ & \\
$u$ & $5.025$  & $0$ & $0$ & $0$ & $+$ & $+$\\
$v,\tilde v$ & $5.025$ & $0$ & $0$ & $\pm 1$ & & $+$
\end{tabular}
\caption{\label{tab:primaries} We identify and measure (the real parts of) several primary fields in the putative CFT for the integrable point at $(\delta,K) = (1,2)$.
Just as chiral primaries with $S \neq 0,N/2$ have an anti-chiral counterpart obtained by reflection (only $\overline \phi$ arises here), also primaries that do not commute with $g_x$ have a counterpart with quantum number $-1$ related by time-reversal symmetry $\Theta$; these are $\tilde\sigma$, $\tilde\sigma'$, and $\tilde v$.
We also resolve charge conjugation $\CC$ for states with $g_x=0$ (these symmetries do not commute), as well as spatial inversion $\II$ in the $0$- and $\pi$-momentum sectors.
The operators above the line are those which we compare with finite-entanglement scaling results for correlations of lattice operators.
}
\end{table}

By finite-size scaling of the energy eigenvalues of highest-weight states we are straightforwardly able to estimate the scaling dimensions of primary operators in the CFT.
Correct normalization of $H$ is evidently very important; to achieve this we follow \citet{milsted2017extraction} and utilize the state related to the stress-energy tensor $T$, which is conserved and has known scaling dimension $\Delta_T = 2$.
$T$ is quasiprimary, related to the vacuum by $\sqrt \frac c2 \ket T = L_{-2} \ket I$ and can thus be readily identified in the $S=2$ sector by calculating $H_{-2}\ket I$.
So $H$ is normalized by setting $\Delta_I = 0$ and $\Delta_T = 2$.
The low-energy spectrum of the model for system size $N=20$ is shown in Fig.~\ref{fig:ed_spec} and the finite-size scaling results are shown in Fig.~\ref{fig:fes_ed_exp}, where they are additionally compared with the finite-entanglement scaling results obtained previously from MPS.

Due to the appearance of the central charge $c$ in the matrix element $\bra T H_{-2} \ket I = \sqrt \frac c2$, we can also compare the finite-size scaling ED results for the central charge with those obtained from MPS.
The finite-size scaling result $c \approx 1.4$ is shown in Fig.~\ref{fig:fss_c}.
While this number is not in agreement with the value obtained previously from scaling with MPS bond dimension, this is not unexpected, as the value of $c$ will drift with system size at a pseudo-critical point, decreasing with increasing system size and eventually reaching $c=0$ at very large sizes.

\section{Discussion}

Motivated by the description of a DQCP in a spin-1/2 chain with rotation symmetry broken to $\Z_2\times\Z_2$, we have probed the nature of a similar transition in a 1d model of local three-level systems forming projective representations of $\Z_3\times\Z_3$.
On one side of the transition is a ferromagnet phase with threefold ground state degeneracy, and on the other a twofold degenerate \VBS phase which preserves onsite symmetries but breaks translation invariance.
This is similar to the $\Z_2\times\Z_2$-symmetric situation, however there an LSM theorem was important in prohibiting an intervening fully symmetric gapped phase; in the present case a featureless phase is allowed.

The above notwithstanding, our studies using an adiabatic protocol for optimized uniform MPS indicate that the phase diagram of the concrete Hamiltonian in Eq.~\eqref{eq:Hz3} does indeed include a direct transition between \zFM and \VBS phases.
Our numerical results are furthermore consistent with a continuous phase transition with symmetry group enhanced to at least $\U1\times\U1\rtimes~\Z_3$.
In addition, the scaling dimensions of the two order parameters involved have nearly the same numerical value, possibly indicating that they are ``unified'' at the transition by a larger emergent symmetry or self-duality.

While we did not obtain a controlled low-energy theory of the transition using either $\Z_3$ domain wall fields or bosonization of the $\U1^2$-symmetric theory (which applies exactly on the lattice along a particular cut through the phase diagram), our numerical results suggest another strategy, by seemingly locating the special point $H^\ast$, Eq.~\eqref{eq:Hstar}, on the phase boundary.
This quantum Hamiltonian is the counterpart to a two-dimensional solvable classical vertex model we term the non-separable integrable NIS model (see Sec.~\ref{sec:integrable}), and through a trick known as transfer matrix inversion one can use the analyticity properties of the eigenvalues to compute exact results about the spectrum.
The surprising result of this method is that $H^\ast$ is gapped, with very long but finite correlation length $\xi = 190878$ lattice spacings.
Such a result is not incompatible with the numerics, which would not distinguish between such approximate conformal symmetry and a truly continuous transition.

The most natural conclusion would seem to be that this DQCP is extremely weakly first order, an intriguing result in light of the status of the \SU2-symmetric DQCP in two dimensions, as discussed in the introduction.
As is true there, the most generic mechanism for generating a hierarchy is through walking, and exact results for $H^\ast$ allow us to write an explicit form for the walking parameter, similar to the case for the self-dual Potts model but with different functional dependence on the continuous tunable parameter, see Eqs.~\eqref{eq:epsilon_sep} and \eqref{eq:epsilon_ns}.
Based on this understanding, we interpret our numerical results as characterizing the (real parts of) the conformal data of the complex CFTs in the walking picture, and we use an ED method to identify some of the light primary fields of these theories.

These developments suggest that the general picture of walking of RG flows is the appropriate way to think about this family of DQCP with $\Z_q\times\Z_q$ symmetry.
In Refs.~\cite{gorbenko2018walking,gorbenko2018walking2} the algebraic equivalence of the Potts model to the six-vertex model plays a crucial role, by allowing through the Coulomb gas formalism many explicit calculations which are then analytically continued into the weakly first-order regime.
The operator algebra of the presented $\Z_q\times\Z_q$ DQCP model is a generalization of the Temperley--Lieb algebra which to our knowledge has not yet demonstrated such equivalences.
A representation theory study of this generalized algebra would be useful in determining whether there are other equivalent models which can illuminate the physics, possibly including a setting for analytic calculations in the ground state.

However, there is also the interesting possibility of qualitatively different walking behaviors arising from the coincidence of the separable and non-separable integrable NIS models at the marginal $q=2$ point.
If these multiple sets of complex CFT fixed points indeed exist in the same parameter space, then for small values of $(q-2)$ one can imagine a rich structure for walking RG flows based on their interactions.
Such a scenario would manifest in crossovers observable in the associated spin chains, and despite the very long length scales involved it is actually possible that quantum Monte Carlo simulations of the explicitly sign-problem-free Hamiltonian in Eq.~\eqref{eq:Hu1} could probe this behavior, along the lines of Refs.~\cite{Desai_2019,roose2020lattice}.
In addition, quantum Monte Carlo studies could be used to test the conjecture about the precise location of the DQCP for $q > 3$, and they could also be used to further examine emergence of the $\U1^2$ symmetry at intermediate scales in the original model Eq.~\eqref{eq:Hz3} with only $\Z_3 \times \Z_3$ symmetry.

Finally, it is not clear what role duality plays in this story.
It seems likely that the successes of duality approaches
in developing descriptions of the DQCP transition in the $\Z_2\times\Z_2$-symmetric model~\cite{jiang2019ising} are special to that model.
However there are some hints in the $\Z_3\times\Z_3$ model: chiefly, the close numerical correspondence of the \zFM and \VBS order parameters is not generally expected and may indicate that the DQCP supports an emergent symmetry or self-dual description.
In addition, the lack of an intervening featureless phase without the help of an anomalous realization of the symmetry on the lattice could be attributable to an emergent anomaly resulting from enhanced symmetry at the transition, which would presumably achieve a ``unification'' of the two order parameters.
It is our hope that further work on the type of one-dimensional model we have studied here will lead to a more complete story of the behaviors of such fixed points in RG space, as well as to a better understanding of how each of these various components contributes to the DQCP phenomenology.

\acknowledgements
We acknowledge helpful conversations with Ashley Milsted, David Simmons-Duffin, Jason Alicea, Yoni BenTov, Cheng-Ju Lin, David Mross, Alex Thomson, Senthil Todari, Christopher White, and Cenke Xu.
This work was supported by National Science Foundation through grants DMR-1619696 and DMR-2001186 (BR and OIM), and by the Institute for Quantum Information and Matter, an NSF Physics Frontiers Center, with support of the Gordon and Betty Moore Foundation (SJ).

\appendix

\section{Review of \texorpdfstring{\SU3}{SU(3)} and \texorpdfstring{\SU3}{SU(3)}-symmetric Hamiltonians} \label{app:su3}

\subsection{Basics of \texorpdfstring{\SU3}{SU(3)}}

The Lie algebra $\mathfrak{su}(3)$ has 8 generators $t^a$, $a=1,\ldots,8$, which in the defining representation $\bm 3$ are represented by the Gell-Mann matrices $\lambda^a$.
We use the alternative convention $T^a = \lambda^a/2$, so the Lie algebra structure constants $f_{abc}$ are determined by \mbox{$[T^a,T^b] = i f_{abc} T^c$}.
The $T^a$ are traceless Hermitian matrices, normalized according to $\mathrm{tr}(T^a T^b) = \frac{1}{2} \delta^{ab}$.
In the conjugate representation $\overline{\bm 3}$ the generators are represented by $\overline{T}^a = -(T^a)^\ast$.

For \SU{q}, $q \geq 2$, one can write a quadratic Casimir invariant
\begin{equation}
C_2 = \sum_a t^a t^a~.
\end{equation}
By construction $C_2$ commutes with all of the $t^a$.
Thus, by Schur's lemma, in an irreducible representation $C_2$ is proportional to the identity.
This operator is familiar from \SU2, where $C_2 = \bm S^2$ and the eigenvalue in an irreducible representation of spin $l$ is $l(l+1)$.
More generally, in a $q$-dimensional representation of \SU{q}, $C_2 = \frac{q^2-1}{2q}$.

\subsection{\texorpdfstring{\SU3}{SU(3)}-invariant Hamiltonians} \label{sec:su3hamiltonians}

In the one-dimensional DQCP with $\Z_2 \times \Z_2$ symmetry studied previously \cite{jiang2019ising,roberts2019deconfined}, a spin Hamiltonian was considered which connects to the solvable Majumdar--Ghosh model.
This ensured the appearance of a phase with VBS order.
That construction generalizes straightforwardly to \SU{q}.
The Majumdar--Ghosh Hamiltonian is the $q=2$ case of
\begin{equation}
H_\mathrm{Cas} = \sum_j \left(C_{2;j,j+1,j+2} - (C_{2;j} + C_{2;j+1} + C_{2;j+2})\right)~,
\label{eq:Hcas}
\end{equation}
where $C_{2;j,j+1,j+2}$ is $C_2$ acting on the tensor product space of three neighboring sites, and $C_2$ is simply a constant on each site individually, as the sites host \SU{q} irreducible representations.
For $q=2$, the fact that the ground states are translation symmetry--breaking products of singlets is a consequence of the irreducible representation decomposition $\bm 2 \otimes \bm 2 = \bm 1 \oplus \bm 3$.
The appearance of the singlet $\bm 1$ is particular to $n=2$; in general, enforcing \SU{q} invariance requires as many single-particle orbitals as internal states.

For $q = 3$, Eq.~\eqref{eq:Hcas} can be used by treating the sites on one sublattice as hosting the conjugate representation $\overline{\bm 3}$.
Then one decomposes $\bm 3 \otimes \overline{\bm 3} = \bm 1 \oplus \bm 8$, so neighboring sites favor an \SU3 singlet.
(A similar statement is true for any $q$, and in fact because $\overline{\bm 2} = \bm 2$ as irreducible representations of \SU{2}, that case is also included.)
The analysis then follows in the same way as for $q=2$.

A local term of $H_\mathrm{Cas}$ is
\begin{equation}
h_{j,j+1,j+2} = \overline T^a_j T^a_{j+1} + T^a_{j+1} \overline T^a_{j+2} + T^a_j T^a_{j+2}~,
\end{equation}
independently of the parity of $j$, as $\overline T^a_j T^a_{j+1} = T^a_j \overline T^a_{j+1}$.
The action of each of these terms can be understood through the action of $C_2$ on tensor products of representations.
Consider
\begin{align}
C_2(\bm 3 \otimes \overline{\bm 3}) &= \sum_a (T^a_j + \overline T^a_{j+1})^2 = 2 \overline T^a_j T^a_{j+1} + \frac 8 3~, \label{eq:c2-1}\\
C_2(\bm 3 \otimes \bm 3) &= \sum_a (T^a_j + T^a_{j+1})^2 = 2 T^a_j T^a_{j+1} + \frac 8 3~. \label{eq:c2-2}
\end{align}
From Eq.~\eqref{eq:c2-1} one learns that $\overline T^a_j T^a_{j+1}$ distinguishes the singlet and the eight-dimensional adjoint representations on sites $j$, $j+1$.
A rank-one projector onto the singlet subspace can thus be written using this term.
Explicitly,
\begin{equation}
\overline T^a_j T^a_{j+1} - \frac 1 6 = -\frac 32 (\Pi_\mathrm{s})_{j,j+1} = -\frac 32 \ket{\psi_\mathrm{s}}\bra{\psi_\mathrm{s}}_{j,j+1}~,
\end{equation}
where $\ket{\psi_\mathrm{s}}_{j,j+1} = \frac{1}{\sqrt{3}} \left(\ket{00}_{j,j+1} + \ket{11}_{j,j+1} + \ket{22}_{j,j+1} \right)$.
Similarly, $\bm 3 \otimes \bm 3 = \overline{\bm 3} \oplus \bm 6$, where $\overline{\bm 3}$ is the antisymmetric subspace and $\bm 6$ the symmetric subspace.
Thus, Eq.~\eqref{eq:c2-2} tells us that
\begin{equation}
T^a_j T^a_{j+1} + \frac 2 3 = \left(\Pi_{\bigvee^2}\right)_{j,j+1}~,
\end{equation}
which is the rank-6 projector onto the symmetric subspace of sites $j$, $j+1$.
(Similar statements apply for general $q$.)
As a result, $H_\text{Cas}$ admits the same arguments that show the ground state manifold of the Majumdar--Ghosh Hamiltonian is spanned by tensor products of \SU{2} singlet dimers, with instead twofold degenerate ground states spanned by products of \SU{q} singlet dimers. 

Conveniently, there is a simpler Hamiltonian than Eq.~\eqref{eq:Hcas} for $q=3$ which exhibits VBS order.
The following nearest-neighbor Hamiltonian was known to \citet{barber1989spectrum} and \citet{affleck1990exact}:
\begin{equation}
H_\text{bQ} = \sum_j \overline T^a_j T^a_{j+1}~.
\label{eq:Hbq}
\end{equation}
This Hamiltonian still respects the full \SU3, and turns out to map exactly to the pure biquadratic \SU{2} spin-1 model.
It is also integrable.
Through its Temperley--Lieb operator algebra this Hamiltonian is related to the XXZ spin-1/2 chain for a particular anisotropy $\Delta = -3/2$ and to the 9-state self-dual Potts model  \cite{barber1989spectrum,sorensen1990correlation}.
The latter equivalence can be seen more directly via a two-step duality procedure which we present in App.~\ref{app:potts_duality}.
Eq.~\eqref{eq:Hbq} turns out to be gapped, with twofold degenerate ground state and finite dimerization order parameter.
Although the ground states are finitely correlated and not a Majumdar--Ghosh-like separable product of dimers, because the ground states respect the \SU3 symmetry we surmise that this Hamiltonian lies in the same phase as $H_\text{Cas}$.
Thus, we consider the local term in $H_\text{bQ}$ to be one favoring a lattice symmetry--breaking but internally symmetric \VBS phase.

\section{MPS for fully symmetric phase and proximate magnetic phase} \label{app:symm_mps}

\subsection{SPT phase with \texorpdfstring{$\Z_3^z \times \Z_3^x$}{Z3xZ3} symmetry}
A gapped fully symmetric ground state is allowed for systems in Eq.~\eqref{eq:Hz3}, and one generically expects to encounter this phase as well.
In fact, this phase has SPT order, since the entanglement spectrum, or boundary states, exhibits degeneracy due to the projective representation.
A simple picture of the phase can be written using an MPS wavefunction of bond dimension three:
\begin{equation}
\ket{\psi_\mathrm{symm}} = \sum_{\{\alpha\}} \Tr\,[\,\cdots A^{\ket{\alpha_j}} A^{\ket{\alpha_{j+1}}} \cdots ]\,\ket{\{\alpha\}}~.
\end{equation}
We choose local tensors to be translationally invariant, so $T_1\ket{\psi_{\mathrm{symm}}}=\ket{\psi_{\mathrm{symm}}}$ automatically.
We also require $A^\ket{\alpha} = (A^\ket{\alpha})^\top$, so that the state is symmetric under inversion.

In order to write a state that is invariant under the action of an onsite symmetry generator $g$, we require that local tensors satisfy the following symmetry condition:
\begin{equation}
A^{\ket{\alpha_j}}=W_{g,j} A^{\ket{\alpha_j}}_g W_{g,j+1}^{-1}~,
\end{equation}
where $A^{\ket{\alpha_j}}_g = g \circ A^{\ket{\alpha_j}}$ and $W_{g,j}$ is an invertible matrix implementing a gauge transformation acting on the left internal leg of the local tensor at site $j$.
The set of $\{W_{g,j}\}_g$ form a projective representation of the symmetry group generated by $\{g\}$.
We choose the virtual legs to index a three-dimensional Hilbert space with basis $\left\{ \ket{0},\ket{1},\ket{2} \right\}$.
The gauge transformations are represented by
\begin{equation}
W_{g,j}=g_{j}\text{ for }g=g_z,g_x,\CC~;~~W_{\Theta,j}=1~.
\end{equation}
The virtual leg $(2k-1,2k)$ hosts the projective representation $[1]$, while the virtual leg $(2k,2k+1)$ carries $[2]$.
Thus, for each tensor one has $[l]+[p]=[r]\bmod 3$, where $[l]$ ($[r]$) labels the projective representation on the left (right) virtual leg, and $[p]$ labels that of the physical leg.

The most general matrices consistent with invariance are
\begin{equation}
A^\ket{0} = \begin{bmatrix} \gamma & 0 & 0 \\ 0 & 0 & \delta \\ 0 & \delta & 0 \end{bmatrix},~A^\ket{1} = \begin{bmatrix} 0 & 0 & \delta \\ 0 & \gamma & 0 \\ \delta & 0 & 0 \end{bmatrix},~A^\ket{2} = \begin{bmatrix} 0 & \delta & 0 \\ \delta & 0 & 0 \\ 0 & 0 & \gamma \end{bmatrix},
\label{eq:z3xz3_spt_mps}
\end{equation}
where $\gamma,\delta \in \R$.
At the special point $\gamma \neq 0$, $\delta=0$ the wavefunction reduces to the ground state of the \zFM phase, and similarly to the ground state of the \xFM phase at $\gamma = \delta \neq 0$.

\subsection{SPT phase with \texorpdfstring{$\U1 \times \U1$}{U1xU1} symmetry}
\label{app:spt_U1xU1}
We now consider the case where $\Z_3^z$ is enlarged to $\U1\times\U1$.
A basis for the legs (physical or virtual) can be labeled by particle numbers $\ket{n_1,n_2}$, which are defined in Eqs.~\eqref{eq:n1def} and \eqref{eq:n2def}.
For the $D=3$ MPS we considered, the physical leg at site $j$ and virtual leg $(j-1,j)$ share the same basis, defined to be
\begin{align}
  \left\{ \ket{0,0}\equiv\ket{0},~\ket{(-1)^j,0}\equiv\ket{1},~\ket{0,(-1)^j}\equiv\ket{2} \right\}~.
  \label{}
\end{align}
The generic form for a local tensor at site $j$ can be represented by a quantum state:
\begin{align}
  \hat{A}_j=\sum (A_j)^{n_1n_2}_{l_1l_2;r_1r_2}\ket{n_1,n_2}_j\otimes\,&\ket{l_1,l_2}_{(j-1,j)} \nonumber \\
  &~~\otimes\bra{r_1,r_2}_{(j,j+1)}.
  \label{eq:tensor_qn_state}
\end{align}

Translation $T_1$ acts as particle-hole symmetry on $\U1 \times \U1$, which relates tensors at even sites $\hat{A}_\text{e}$ and those at odd sites $\hat{A}_\text{o}$ via
\begin{align}
  (A_\text{o})^{n_1n_2}_{l_1l_2;r_1r_2} = (A_\text{e})^{-n_1,-n_2}_{-l_1,-l_2;-r_1,-r_2} ~.
  \label{eq:tensor_t1_cons}
\end{align}

For a $\U1\times\U1$ symmetric MPS, $\hat{A}_j$ in Eq.~\eqref{eq:tensor_qn_state} should satisfy the particle number conservation condition
\begin{align}
  n_a+l_a=q_a+r_a~,\quad\text{where } a=1,2 ~.
  \label{eq:tensor_no_cons}
\end{align}
Here, $q_a$ is a site-dependent constant.
On a periodic chain, this state has definite total particle numbers $\N_a \equiv \sum_j n_{a,j} = \sum_j q_{a,j}$, $a=1,2$.

By construction, a generic MPS in Eq.~\eqref{eq:z3xz3_spt_mps} breaks $\U1\times\U1$ symmetry to $\Z_3^z$.
However, $\U1\times\U1$ symmetry can be restored by setting $\gamma=0$.
Indeed, in this case the local tensors can be written
\begin{widetext}
\begin{align}
  \hat{A}_\text{e} &= \ket{0,0}\otimes\big( \ket{1,0}\bra{0,-1} + \ket{0,1}\bra{-1,0} \big) + \ket{1,0}\otimes\big( \ket{0,1}\bra{0,0} +  \ket{0,0}\bra{0,-1} \big) + \ket{0,1}\otimes\big( \ket{0,0}\bra{-1,0} + \ket{1,0}\bra{0,0} \big)~,\\
  \hat{A}_\text{o} &= \ket{0,0}\otimes\big( \ket{-1,0}\bra{0,1} + \ket{0,-1}\bra{1,0} \big) + \ket{-1,0}\otimes\big( \ket{0,-1}\bra{0,0} + \ket{0,0}\bra{0,1} \big) + \ket{0,-1}\otimes\big( \ket{0,0}\bra{1,0} + \ket{-1,0}\bra{0,0} \big)~,
 \label{}
\end{align}
\end{widetext}
where we have dropped the overall amplitude $\delta$.
One can check that these tensors indeed satisfy Eq.~\eqref{eq:tensor_no_cons} with $q_a = 1$ ($-1$) for even (odd) sites.
The other symmetries of the model, $\II$, $\Theta$, $g_x$ and $\CC$, are also preserved by this MPS.

However, for the purpose of obtaining an MPS beyond the $D=3$ case we can work out the symmetry constraints on $A_j$.
Constraints from $T_1$ and $\U1\times\U1$ are already listed in Eqs.~\eqref{eq:tensor_t1_cons} and \eqref{eq:tensor_no_cons}.
Time reversal $\Theta$ simply requires all tensor entries to be real numbers.

To be consistent with $\U1 \times \U1$ symmetry in Eq.~\eqref{eq:tensor_no_cons}, inversion $\II$ acts with an additional particle-hole symmetry on the virtual legs, imposing the following constraint:
\begin{equation}
(A_j)_{l_1l_2;r_1r_2}^{n_1n_2} = (A_j)_{-r_1,-r_2;-l_1,-l_2}^{n_1n_2}~.
\label{eq:tensor_inv_cons}
\end{equation}
$\CC$ interchanges particles between the two species, thus
\begin{equation}
(A_j)^{n_1n_2}_{l_1l_2;r_1r_2} = (A_j)^{n_2n_1}_{l_2l_1;r_2r_1} ~.
\label{eq:tensor_cc_cons}
\end{equation}
On the physical leg at site $j$, $g_x$ maps $\ket{n_1,n_2}_j$ to $\ket{(-1)^j-n_1-n_2,n_1}_j$.
On the left virtual leg $(j-1,j)$, the action of $g_x$ is the same:
\begin{equation}
g_x: \ket{l_1,l_2}_{(j-1,j)}\to \ket{(-1)^j-l_1-l_2,l_1}_{(j-1,j)}~,
\label{}
\end{equation}
while on the right legs the fact that these are contracted with the left legs on the next tensor fixes the transformation to be
\begin{equation}
g_x: \bra{r_1,r_2}_{(j,j+1)}\to \bra{(-1)^{j+1}-r_1-r_2,r_1}_{(j,j+1)} ~. \nonumber
\label{}
\end{equation}
Thus, $g_x$ imposes the constraint
\begin{equation}
(A_j)^{n_1n_2}_{l_1l_2;r_1,r_2} = (A_j)^{(-1)^j-n_1-n_2,n_1}_{(-1)^j-l_1-l_2,l_1;(-1)^{j+1}-r_1-r_2,r_1}
\label{eq:tensor_gx_cons}
\end{equation}
In summary, to construct a fully symmetric MPS with site tensor $\hat{A}_j$ defined in Eq.~\eqref{eq:tensor_qn_state}, tensor entries $(A_{j})_{l_1l_2;r_1r_2}^{n_1n_2}$ should be real numbers, and satisfy the symmetry conditions Eq.~\eqref{eq:tensor_t1_cons}, \eqref{eq:tensor_no_cons}, \eqref{eq:tensor_inv_cons}, \eqref{eq:tensor_cc_cons}, and \eqref{eq:tensor_gx_cons}.

\subsection{Bond-centered magnetic order phase} \label{app:phase_connect_zFM_spt}

In this part, we present an MPS construction for the bond-centered magnetic order phase, which is the intermediate phase smoothly connecting the \zFM and SPT phases in the classical phase diagram, as shown in Fig.~\ref{fig:luttinger_classical_phase_diagram}.
Although it is a spontaneously symmetry breaking phase with six-fold ground state degeneracy, its ground states cannot be represented by direct product states.

We start from the MPS representation of the SPT phase with $\U1\times\U1$ symmetry.
As discussed in the last part, this MPS is constructed from a site tensor $A$ in Eq.~\eqref{eq:z3xz3_spt_mps} with $\gamma=0$.
We can represent $A$ as quantum state as
\begin{equation}
\hat{A}=\sum_{a=0}^2\ket{a}\otimes(\ket{a-1}\bra{a+1}+\ket{a+1}\bra{a-1})~.
\label{eq:bond_magnetic_order_site_tensor}
\end{equation}
Let us insert additional bond tensors $B_{j,j+1}$ sitting between sites $j$ and $j+1$.
For the SPT phase, $B_{j,j+1}$ is the identity matrix, whose quantum state representation is
\begin{equation}
\hat{B}_{j,j+1}=\sum_{a=0}^2 \ket{a}\bra{a}
\end{equation}

We now break some symmetry by introducing a parameter $\kappa$ into the bond tensors:
\begin{align}
    \hat{B}_{2k-1,2k}=(1-\kappa)\ket{0}\bra{0}+(1-\kappa)\ket{1}\bra{1}+(1+\kappa)\ket{2}\bra{2}~,\notag\\
    \hat{B}_{2k,2k+1}=(1-\kappa)\ket{0}\bra{0}+(1+\kappa)\ket{1}\bra{1}+(1-\kappa)\ket{2}\bra{2}~,
    \label{eq:bond_magnetic_order_bond_tensors}
\end{align}
where $0 \leq \kappa \leq 1$.
We leave the site tensors unchanged.
When $\kappa=0$, we recover the SPT state.
When $\kappa=1$, $\hat{B}_{2k-1,2k}=2\ket{2}\bra{2}$ and $\hat{B}_{2k,2k+1}=2\ket{1}\bra{1}$, and by contracting all virtual legs, we get a \zFM state $\bigotimes_j\ket{0}_j$ (up to a constant).
Thus, this state indeed smoothly connects between the SPT and \zFM phases.

We now analyze symmetry properties for the state with $0<\kappa<1$.
The action of onsite symmetries on virtual legs is discussed in App.~\ref{app:spt_U1xU1}.
It is straightforward to see that this state preserves $\U1\times\U1$ symmetry and breaks $g_x$, $\CC$, $T_1$, and $\II$ symmetries.
In fact, $T_1$, $\II$, and $\CC$ act in the same way on this MPS, producing a state with even and odd bond tensors in Eq.~\eqref{eq:bond_magnetic_order_bond_tensors} interchanged:
\begin{align}
    \hat{B}_{2k-1,2k}=(1-\kappa)\ket{0}\bra{0}+(1+\kappa)\ket{1}\bra{1}+(1-\kappa)\ket{2}\bra{2}~,\notag\\
    \hat{B}_{2k,2k+1}=(1-\kappa)\ket{0}\bra{0}+(1-\kappa)\ket{1}\bra{1}+(1+\kappa)\ket{2}\bra{2}~,
\end{align}
We note that this pair of MPS share the same symmetry properties as states labeled by $(\pi/6\pm\upsilon,\pi/6\mp\upsilon)$ in Eq.~\eqref{eq:bond_order_phase}.
The MPS representation of the other two pairs of states in Eq.~\eqref{eq:bond_order_phase} can be generated by the action of $g_x$.
Note that site tensors are invariant under $g_x$ symmetry, and are given by Eq.~\eqref{eq:bond_magnetic_order_site_tensor}. 
Bond tensors for the MPS states corresponding to $(\pi/6\mp\upsilon,-\pi/3)$ are
\begin{align}
    \hat{B}_{2k-1,2k}=(1\pm\kappa)\ket{0}\bra{0}+(1-\kappa)\ket{1}\bra{1}+(1\mp\kappa)\ket{2}\bra{2}~,\notag\\
    \hat{B}_{2k,2k+1}=(1\mp\kappa)\ket{0}\bra{0}+(1-\kappa)\ket{1}\bra{1}+(1\pm\kappa)\ket{2}\bra{2}~,
\end{align}
and the bond tensors for states corresponding to $(-\pi/3,\pi/6\pm\upsilon)$ are
\begin{align}
    \hat{B}_{2k-1,2k}=(1\mp\kappa)\ket{0}\bra{0}+(1\pm\kappa)\ket{1}\bra{1}+(1-\kappa)\ket{2}\bra{2}~,\notag\\
    \hat{B}_{2k,2k+1}=(1\pm\kappa)\ket{0}\bra{0}+(1\mp\kappa)\ket{1}\bra{1}+(1-\kappa)\ket{2}\bra{2}~.
\end{align}

\section{Domain wall duality mapping with \texorpdfstring{$\Z_3$}{Z3} gauge field} \label{app:z3_dual}

In this section we present the more precisely defined version of the duality mapping to domain walls on a periodic chain, which appear as matter fields on the dual lattice coupled to a $\Z_3$ gauge field.
The purpose of the gauge field is essentially for bookkeeping, as it does not have its own dynamics.
Instead, it will account for the differing global properties of the phases, the most important example in our case being ground state degeneracy.

In addition to the domain wall variables $\XX_{j+1/2}$, $\ZZ_{j+1/2}$ which live on the sites of the dual lattice, we place gauge degrees of freedom $\rho^x_j$, $\rho^z_j$ which form a [1] projective representation of $\Z_3 \times \Z_3$ on the links of the dual lattice (equivalently, on the sites of the primal lattice).
The duality mapping is then given by
\begin{align}
\XX_{j+1/2} &= Z^\dag_j Z_{j+1}~, \\
\ZZ^\dag_{j-1/2} \; \rho^{z\dag}_j \; \ZZ_{j+1/2} &= X_j~, \\
\rho^x_j &= Z_j~.
\end{align}
The physical Hilbert space satisfies the gauge constraint
\begin{equation}
\XX_{j+1/2} = \rho^{x\dag}_j \rho^x_{j+1}~.
\end{equation}
The Hamiltonian Eq.~\eqref{eq:Hz3} translates to
\begin{align}
\wt{H} = -\sum_j &\Big[(J^x \ZZ^\dag_{j-1/2} \rho^{z\dag}_j \rho^{z\dag}_{j+1} \ZZ_{j+3/2} + J^z \XX_{j+1/2} + \Hc) \nonumber \\
&\quad +K (1 + \ZZ^\dag_{j-1/2} \rho^{z\dag}_j \rho^{z\dag}_{j+1} \ZZ_{j+3/2} + \Hc) \nonumber \\
&\qquad\qquad\qquad\times(1 + \XX_{j+1/2} + \Hc)\Big]~.
\end{align}
Using the dictionary above, and requiring equality to hold only in the physical sector, we can also rewrite the symmetry generators as
\begin{equation}
g_x = \prod_j \rho^{z\dag}_j~,~~g_z = \prod_k \XX_{2k+1/2} = \prod_k \rho^{x\dag}_{2k} \rho^x_{2k+1}~,
\end{equation}
which are exact on a periodic system.
Now one obtains the duality mapping presented in Sec.~\ref{sec:duality} by fixing the gauge $\rho^z_j = 1$.
The action of the symmetries on the gauge variables is given by
\begin{align}
g_x:~~&(\rho^x_j,\rho^z_j)\mapsto(\omega^{-1}\rho^x_j,\rho^z_j)~,\\
g_z:~~&(\rho^x_j,\rho^z_j)\mapsto(\rho^x_j,\omega^{1-2p_j}\rho^z_j)~,\\
\Theta:~~&(\rho^x_j,\rho^z_j)\mapsto(\rho^{x\dag}_j,\rho^z_j)~,\\
\CC:~~&(\rho^x_j,\rho^z_j)\mapsto(\rho^{x\dag}_j,\rho^{z\dag}_j)~,\\
T_1:~~&(\rho^x_j,\rho^z_j)\mapsto(\rho^x_{j+1},\rho^z_{j+1})~,\\
\II:~~&(\rho^x_j,\rho^z_j)\mapsto(\rho^x_{-j},\rho^z_{-j})~.
\end{align}
Importantly, $g_x$ acts nontrivially in this formulation.
As in the main text, we designate the ``even'' and ``odd'' sublattices of the dual lattice as locations $2k+1/2$ and $2k+3/2$, $k \in \mathbb{Z}$, respectively.

We refer to this theory as having a $\Z_3^\rho$ gauge symmetry.
Briefly, the pure gauge theory with physical constraint $\rho^{x\dag}_j \rho^x_{j+1} = 1$ comprises three sectors, specified by $\rho^x_j = \omega^r$ for $r=0,1,2$.
These sectors are related by the symmetry generator $\prod_j \rho^{z\dag}_j = g_x$, which is a symmetry of the Hamiltonian.
Thus the appropriate sectors of the gauge symmetry are the linear combinations respecting $g_x$, namely with definite flux $\prod_j \rho^z_j$ taking values $1$, $\omega$, or $\omega^2$.
The instanton operator adding $\Z_3^\rho$ flux is $\rho^x_j$, which is indeed seen to transform nontrivially under $g_x$.

\subsection{Symmetry-breaking phases from the dual perspective}

We can now revisit the phases described in Sec.~\ref{sec:duality}.
Consider first the case in which domain walls are gapped, so the low-energy properties are determined simply by the gauge sector.
In this case we have $\lrangle{\ZZ_{j+1/2}} = 0$; this pattern is energetically favored in our model for $J^z$ dominant.
Because the instanton operator is not included in the Hamiltonian the three gauge flux sectors do not mix.
From a formal perspective where we integrate out the gapped matter field $\ZZ$, the three states with different flux $\prod_j \rho_j^z$ can obtain slightly different energies but the energy splitting is exponentially small in the chain length.
This corresponds to spontaneously breaking $g_x$ and accounts for the threefold degeneracy of the ground state in the \zFM phase.

The domain wall condensate having $\lrangle{\ZZ_\odd} \neq 0$, $\lrangle{\ZZ_\even} \neq 0$ leads to a Higgs phase of the gauge field.
Minimizing the energy of the $J^x$ terms, it must be that $\prod_j \rho_j^z = 1$; {\it i.e.}, a unique gauge flux is selected and hence the $g_x$ symmetry is respected.
Solving for classical ground states, there are three gauge-inequivalent solutions with this flux, with representative states $\rho_j^z = 1, \ZZ_\odd = 1, \ZZ_\even = \omega^p$ everywhere on the chain, with $p = 0, \pm 1$.
These solutions are distinguished by gauge-invariant observables $\ZZ_{j-1/2}^\dagger \rho_j^{z\dagger} \ZZ_{j+1/2}$, which are the same as the original physical $X_j$ variables, and the resulting three different patterns in these correspond to the three \xFM ground states in Eq.~\eqref{eq:BxFM}.
We can thus see from the matter fields that $g_z$ is broken but spatial symmetries are respected.
All of these cases, which are favored at large values of $J^x$, make up the \xFM phase with threefold degeneracy.
In more schematic terms, in the absence of the gauge field we would have separate $\mathbb{Z}_3$ symmetry associated with each  of the ``even'' and ``odd'' sublattices of the dual lattice.
Simultaneous condensation $\lrangle{\ZZ_\odd} \neq 0$, $\lrangle{\ZZ_\even} \neq 0$ would then produce nine ground states. 
However, the presence of the dual gauge field and the Higgs mechanism will reduce the number of true ground states down to three as discussed above.

We can also consider a condensate $\lrangle{\ZZ_\odd} \neq 0$ and $\lrangle{\ZZ_\even} = 0$, or vice versa.
As was the case in the \xFM phase, the Higgs mechanism here restores the $g_x$ symmetry by selecting a unique flux sector $\prod_j \rho_j^z = 1$,
but in contrast to the previous case, $g_z$ and other internal symmetries are respected as well.
(Schematically, the naive three-fold degeneracy from condensing $\ZZ$ on one sublattice is reduced down to one by the Higgs mechanism.)
The state does break a $\Z_2$ translation symmetry however, and therefore is identified as the \VBS phase.
It is not evident from this analysis that this phase is energetically favored at large $K$ in our model, but ample evidence of this fact is obtained from other sources.

\subsection{SPT phase from the dual perspective}

To obtain a fully symmetric phase, we condense a bound state of a domain wall on the odd sublattice and a domain wall on the even sublattice: schematically, $\lrangle{\ZZ_{\text{odd}} \ZZ_\text{even}} \neq 0$ while $\lrangle{\ZZ_\text{odd}} = \lrangle{\ZZ_\text{even}} = 0$.
The $g_x$ symmetry is restored because this bound state carries unit dual gauge charge:  Indeed, keeping track of only the dual gauge charge, we have schematically $\ZZ^2 \sim \ZZ^{-1}$ (note that it is crucial that we have $\mathbb{Z}_N$ gauge field with odd $N$).
Hence, the $\ZZ_\text{odd} \ZZ_\text{even}$ condensate completely Higgses out the dual gauge field $\rho$, which corresponds to the presence of the $g_x$ symmetry.
Since translation interchanges $\ZZ_\text{odd}$ and $\ZZ_\text{even}$, this condensate clearly preserves this symmetry.
Under $g_z$ action, $\ZZ_\text{odd} \ZZ_\text{even}$ obtains a phase factor $\omega^2$; however, this is related to the fact that this schematic object is not gauge-invariant and the phase factor can be removed by a gauge transformation.
Any gauge-invariant local operator with non-zero expectation value will respect the $g_z$ symmetry.
Thus, we obtain a fully symmetric phase.

Another perspective on this condensate is that we condense bound states of a domain wall field in the $g_x$-symmetry-breaking order ({\it i.e.}, $\ZZ$ field) and a $g_z$ charge field ({\it i.e.}, $X$ field).
Indeed,
$\ZZ_{j-1/2} \ZZ_{j+1/2} = \ZZ_{j-1/2}^2 \rho_j^z X_j \sim \ZZ_{j-1/2}^\dagger X_j$ (fixing the gauge $\rho_j^z=1$).
We expect that condensation of bound states of domain walls and charges leads to a non-trivial SPT phase.

\section{Duality of \texorpdfstring{$q$}{q}-state separable model and \texorpdfstring{$q^2$}{q2}-state Potts model and generalization to non-separable model} \label{app:potts_duality}

In this Appendix, we perform a two-step duality that connects the $q$-state separable integrable model and $Q_\text{Potts} = q^2$-state Potts model.
We will also follow the non-separable integrable model under the same mapping.
The treatment here is in the Hamiltonian language and can be carried out for any integer $q$.

We begin with a $q$-state generalization of the $\U1^2$-symmetric $q = 3$ model from the main text.
Consider the Hamiltonian
\begin{align}
H = -\sum_j \Bigg[ & J_z \sum_{\ell=0}^{q-1} \left(Z_j^\dagger Z_{j+1} \right)^\ell \\
& + K \sum_{\ell=0}^{q-1} \left(X_j X_{j+1} \right)^\ell ~\sum_{\ell=0}^{q-1} \left(Z_j^\dagger Z_{j+1} \right)^\ell \Bigg] ~. \nonumber
\end{align}
For $q=3$ this reduces to the model in the main text, up to an additive constant. 
For general $q$ the terms in the Hamiltonian have a simple form in bra-ket notation (see also Eq.~\eqref{eq:Hu1}):
\begin{align*}
& \sum_{\ell=0}^{q-1} \left(Z_j^\dagger Z_{j+1} \right)^\ell = q \sum_\alpha \ket{\alpha,\alpha} \bra{\alpha,\alpha}_{j,j+1} ~, \\
& \sum_{\ell=0}^{q-1} \left(X_j X_{j+1} \right)^\ell ~\sum_{\ell=0}^{q-1} \left(Z_j^\dagger Z_{j+1} \right)^\ell\!= q \sum_{\alpha,\beta} \ket{\beta,\beta} \bra{\alpha,\alpha}_{j,j+1} ~,
\end{align*}
from which it is easy to see that the model has continuous $\U1^{q-1}$ symmetry as well as $S_q$ permutation symmetry.
It has a trivial solvable point $J_z > 0, K = 0$ inside the \zFM phase as well as two nontrivial integrable points: 
$J_z = 0, K > 0$ which is inside the VBS phase, and $J_z = K(q-2) > 0$ which we propose is at the transition between the \zFM and \VBS phases.

We first perform a formal duality transformation which is a straightforward $q$-state generalization of the one in the main text:
\begin{align}
& X_j = \ZZ_{j-1/2}^\dagger \ZZ_{j+1/2} ~, \\
& Z_j^\dagger Z_{j+1} = \XX_{j+1/2} ~.
\end{align}
(For simplicity here and below, we do not exhibit dual gauge fields which would be necessary to account for global aspects in a periodic chain.) 
The dual Hamiltonian reads
\begin{align}
\wt{H} = -\sum_j \Bigg[ & J_z \sum_{\ell=0}^{q-1} \left(\XX_{j+1/2} \right)^\ell \\
& + K \sum_{\ell=0}^{q-1} \left(\ZZ_{j-1/2}^\dagger \ZZ_{j+3/2} \right)^\ell ~\sum_{\ell=0}^{q-1} \left(\XX_{j+1/2} \right)^\ell \Bigg] . \nonumber
\end{align}
Similarly to the main text, this can be viewed as two individually Potts-symmetric $q$-state systems residing on the ``even'' and ``odd'' sublattices of the dual lattice (locations $2k+1/2$ and $2k+3/2$, $k \in \mathbb{Z}$, respectively).
The two systems have energy-energy coupling between them.
In these variables, the \zFM phase occurs when both $\ZZ_{2k+1/2}$ and $\ZZ_{2k+3/2}$ are gapped.
On the other hand, the \VBS phase occurs when only one species orders but not the other, which breaks the translation symmetry.

Let us now maintain the even sublattice variables $(\ZZ_{2k+1/2}, \XX_{2k+1/2})$ and perform the above duality transformation on the odd sublattice variables $(\ZZ_{2k+3/2}, \XX_{2k+3/2})$, treating this system as a 1d chain:
\begin{align}
& \XX_{2k+3/2} = \dwt{Z}_{2k+1/2}^\dagger \dwt{Z}_{2k+5/2} ~, \\
& \ZZ_{2k-1/2}^\dagger \ZZ_{2k+3/2} = \dwt{X}_{2k+1/2} ~.
\end{align}
Note that the variables dual to $(\ZZ_{2k+3/2}, \XX_{2k+3/2})$ reside at the same locations as the even sublattice variables $(\ZZ_{2k+1/2}, \XX_{2k+1/2})$, as indicated by the location indices of $(\dwt{Z}_{2k+1/2}, \dwt{X}_{2k+1/2})$.
After this transformation, the Hamiltonian reads:
\begin{widetext}
\begin{align}
\label{eq:dbtildeH}
\dwt{H} = -\sum_{k \in \Z} \Bigg[& J_z \sum_{\ell=0}^{q-1} \left(\XX_{2k+1/2} \right)^\ell 
+ J_z \sum_{\ell=0}^{q-1} \left(\dwt{Z}_{2k+1/2}^\dagger \dwt{Z}_{2k+5/2} \right)^\ell \\
& + K \sum_{\ell=0}^{q-1} \left(\dwt{X}_{2k+1/2} \right)^\ell ~\sum_{\ell=0}^{q-1} \left(\XX_{2k+1/2} \right)^\ell
+ K \sum_{\ell=0}^{q-1} \left(\ZZ_{2k+1/2}^\dagger \ZZ_{2k+5/2} \right)^\ell ~\sum_{\ell=0}^{q-1}
\left(\dwt{Z}_{2k+1/2}^\dagger \dwt{Z}_{2k+5/2} \right)^\ell
\Bigg] ~.
\end{align}
\end{widetext}
In these variables, the \zFM phase corresponds to gapped $\ZZ_{2k+1/2}$ variables and condensed $\dwt{Z}_{2k+1/2}$ variables.
On the other hand, the \VBS phase corresponds to either both $\ZZ_{2k+1/2}$ and $\dwt{Z}_{2k+1/2}$ being gapped or both condensed.

We can combine the tilded and double-tilded variables on each site $2k+1/2$ to form a $q^2$-state variable, $\ket{A}_{2k+1/2} \equiv \ket{\wt{\alpha}}_{2k+1/2} \otimes \ket{\dwt{\alpha}}_{2k+1/2}$, $\wt{\alpha}, \dwt{\alpha} = 1, \dots, q$.
The $K$ terms become precisely the on-site and inter-site quantum Potts terms for these $Q_\text{Potts}=q^2$-state variables:
\begin{widetext}
\begin{align}
& \sum_{\ell=0}^{q-1} \left(\XX_{2k+1/2} \right)^\ell ~\sum_{\ell=0}^{q-1} \left(\dwt{X}_{2k+1/2} \right)^\ell
= \sum_{\wt{\alpha},\wt{\beta}} \ket{\wt{\beta}} \bra{\wt{\alpha}}_{2k+1/2} \otimes \sum_{\footnotesize\dwt{\alpha},\dwt{\beta}} \ket{\dwt{\beta}} \bra{\dwt{\alpha}}_{2k+1/2} 
= \sum_{A,B} \ket{B} \bra{A}_{2k+1/2} \equiv \sum_{\ell=0}^{q^2-1} \left(\mathcal{X}_{2k+1/2} \right)^\ell~,
\end{align}
\begin{align}
\sum_{\ell=0}^{q-1} \left(\ZZ_{2k+1/2}^\dagger \ZZ_{2k+5/2} \right)^\ell ~\sum_{\ell=0}^{q-1}
\left(\dwt{Z}_{2k+1/2}^\dagger \dwt{Z}_{2k+5/2} \right)^\ell 
&= q \sum_{\wt{\alpha}} \ket{\wt{\alpha},\wt{\alpha}} \bra{\wt{\alpha},\wt{\alpha}}_{2k+1/2,2k+5/2} \otimes q \sum_{\footnotesize\dwt{\alpha}} \ket{\dwt{\alpha},\dwt{\alpha}} \bra{\dwt{\alpha},\dwt{\alpha}}_{2k+1/2,2k+5/2} \nonumber \\
& = q^2 \sum_A \ket{A,A} \bra{A,A}_{2k+1/2,2k+5/2} \equiv \sum_{\ell=0}^{q^2-1} \left(\mathcal{Z}_{2k+1/2}^\dagger \mathcal{Z}_{2k+5/2} \right)^\ell ~,
\end{align}
\end{widetext}
where we have introduced standard operators $\mathcal{Z}_{2k+1/2}, \mathcal{X}_{2k+1/2}$ in the $Q_\text{Potts} = q^2$-state Hilbert space on each site $2k+1/2$.
Thus, in the absence of the $J^z$ term we indeed obtain the self-dual $q^2$-state Potts model on the ``even'' sublattice of the dual lattice.
This type of equivalence of the integrable model $H[J_x=0,J_z=0, K]$ to the self-dual $q^2$-state Potts model has been well known at least since Refs.~\cite{barber1989spectrum,affleck1990exact} where it was argued by comparing the Temperley--Lieb operator algebras in the two models.
This is the quantum version of the equivalence between the classical separable integrable NIS and classical $q^2$-state Potts models mentioned in Sec.~\ref{sec:integrable}.
By examining the origins of the two $K$ terms in Eq.~\eqref{eq:dbtildeH}, it is also easy to see that staggering bond couplings in the original model corresponds to moving off self-duality in the Potts model.

The derivation here is of some interest in that it clearly demonstrates a non-local relation between the two models and also allows one to formulate the precise relation on periodic chains by carefully including the gauge fields appearing in the dualities to keep track of the global aspects, which for the sake of simplicity we did not include.
Of particular interest to us is that we can also write the $J^z$ terms, which from Eq.~\eqref{eq:dbtildeH} are
\begin{align}
& \sum_{\ell=0}^{q-1} \left(\XX_{2k+1/2} \right)^\ell = \sum_{\ell=0}^{q-1} \left(\mathcal{X}_{2k+1/2} \right)^{\ell \cdot q} ~, \\ 
& \sum_{\ell=0}^{q-1} \left(\dwt{Z}_{2k+1/2}^\dagger \dwt{Z}_{2k+5/2} \right)^\ell =
\sum_{\ell=0}^{q-1} \left(\mathcal{Z}_{2k+1/2}^\dagger \mathcal{Z}_{2k+5/2} \right)^{\ell \cdot q} ~.
\end{align}
Note that the powers of operators summed on the right hand side are $\ell \cdot q$, which appear in the convention of the following ordering of the $q^2$ states $\ket{A} = \ket{\wt{\alpha}} \otimes \ket{\dwt{\alpha}}$:
\begin{equation}
    A = (\wt{\alpha} - 1) \cdot q + \dwt{\alpha} ~,
\end{equation}
$\wt{\alpha}, \dwt{\alpha} = 1, \dots, q; A = 1, \dots, q^2$.
We can now see that the $q^2$-state model remains self-dual also in the presence of the $J_z$ term, which however breaks the formal symmetry in these variables from $S_{q^2}$ down to $S_q \times S_q$, as is clear from Eq.~\eqref{eq:dbtildeH}.
Unfortunately, this formulation does not appear to inform us why $J_z = K(q-2)$ places the model precisely at the transition between the \zFM and \VBS phases, which in the $\ZZ_{2k+1/2}$ and $\dwt{Z}_{2k+1/2}$ variables are described after Eq.~\eqref{eq:dbtildeH}.
In the $q^2$-state Potts variables $\mathcal{Z}_{2k+1/2}$, the \VBS phase corresponds to the first-order coexistence of the standard disordered and ordered Potts phases, while the \zFM phase corresponds to a specific partial order.
In this language, $J_z = K(q-2)$ appears to correspond to a special multi-critical point, and we are hopeful that this information may be useful for future elucidation of this transition.

\bibliography{refs}

\end{document}